# RIS-Enabled UAV Communications and Sensing: Opportunities, Challenges, and Key Technologies

Yajun Zhao, *Member, IEEE*, Mengnan Jian, *Member, IEEE*, Yifei Yuan, *Fellow, IEEE*

*Abstract*—Unmanned Aerial Vehicles (UAVs) play a pivotal role in the emerging low-altitude economy. However, they face significant challenges in achieving reliable network coverage during transit operations. This paper provides an in-depth investigation into the characteristics and challenges of communication networks tailored for UAVs. First, we outline typical operational scenarios, traffic patterns, and a dual-layer heterogeneous network topology. This topology is essential for enabling three-dimensional continuous coverage and ensuring seamless network coexistence between UAVs and other network entities. Moreover, the paper delves into the channel characteristics and specific challenges faced by UAV Integrated Sensing and Communication (ISAC) networks. It highlights the limitations of traditional Active Phased Array Antenna (APAA)-based networks, particularly regarding cost, complexity, and site deployment constraints. We then introduce Reconfigurable Intelligent Surface (RIS)-assisted networks as a promising solution for enhancing UAV signal coverage. The key technical features of RIS are discussed, including design principles, antenna tilt configurations, new beam types, and beam tracking mechanisms. In addition, we examine the impact of high-frequency bands and their absorption peaks on signal attenuation. The paper further explores network architecture designs aimed at improving UAV signal coverage, facilitating network coexistence, and supporting RIS-enhanced UAV sensing. Field trial results evaluating the effectiveness of RIS in improving UAV coverage are presented. Finally, we outline future technological trends and highlight potential advancements to further optimize UAV communication systems. We also emphasize the importance of engineering implementation and standardization efforts in RIS-based UAV-ISAC networks.

## I. INTRODUCTION

ACCORDING to a report by MarketsandMarkets, the Unmanned Aerial Vehicle (UAV) market is projected to grow from USD 30.2 billion in 2024 to USD 48.5 billion by 2029, reflecting a compound annual growth rate (CAGR) of 9.9% during this period[1]. This growth is primarily driven by the continuous integration and upgrading of drones across diverse applications. Technological advancements, including enhanced power systems, sophisticated sensing technologies, and increased autonomy, have significantly expanded UAV capabilities. Furthermore, the evolution of regulations facilitating the deployment of drones in civilian and commercial sectors has notably contributed to market expansion. The rising demand from industries such as delivery services, environmental surveillance, and infrastructure monitoring plays a crucial role in this trajectory. The convergence of UAV technology with emerging digital innovations continues to propel the industry's growth. However, the coverage issues of UAV Integrated Sensing and Communication (UAV-ISAC) networks are technical bottlenecks for the large-scale deployment of UAVs.

### A. Brief Review of UAV, ISAC and RIS

#### 1) Unmanned Aerial Vehicle (UAV)

UAVs can be classified into various categories based on factors such as weight, size, wing configuration, and flight time, including small/micro-UAVs and large UAVs, as well as fixed-wing and rotor-wing types [1]. Each UAV type possesses distinct characteristics, making them suited for different application scenarios. A comprehensive overview of UAV classifications and applications was provided in [2].

Wireless communication is one of the key technologies for unlocking the full potential of UAVs. As such, it has recently garnered unprecedented attention [3]-[5]. Currently, research into UAV networks has surged. Several studies have focused on developing a comprehensive understanding of UAV communication models. These investigations also address various aspects related to UAV applications, characteristics, challenges, and unresolved issues [6]-[8]. Additionally, some surveys propose solutions to specific requirements, such as security concerns, media access control protocols, quality of service (QoS), and routing protocols. The study in [6] systematically reviews on the use of UAV in wireless networks, offering a detailed examination of the fundamental trade-offs and significant challenges in UAV wireless networks.

Recent advancements have led to two promising avenues in UAV communications: UAV-assisted cellular networks and cellular-connected UAVs. Consequently, the integration of UAVs into cellular networks is recognized as a synergistic innovation beneficial to both UAV-related industries and cellular service providers. In this context, UAV communications can be categorized based on the role of the UAV within the network: 1) UAVs functioning as base stations (BS) (UAV-BS) [9] or relays to enhance signal

(Corresponding author: Yajun Zhao).

Yajun Zhao is with ZTE Corporation, Beijing, CO 100029, China. (e-mail: zhao.yajun1@zte.com.cn).

Mengnan Jian is with ZTE Corporation, Beijing, CO 100029, China. (e-mail: jian.mengnan@zte.com.cn).

Yifei Yuan is with the Future Research Lab, China Mobile Research Institute, Beijing, CO 100053, China, (e-mail: yuanyifei@chinamobile.com).

[1] https://www.researchandmarkets.com/r/3t8qve.

coverage for ground communication or to facilitate terminal and target sensing, and 2) UAVs operating as communication terminals/user equipment (UE) (UAV-UE) [10] or sensing targets that depend on terrestrial networks for coverage, particularly in low-altitude scenarios. Within these frameworks, UAVs are integrated into cellular systems, fulfilling dual roles as both aerial communication platforms and users. This integration not only opens up a wide range of new business opportunities but also enhances the performance of three-dimensional (3D) wireless networks [11].

2) **Integrated Sensing and Communication (ISAC)**

In September 2023, the International Telecommunication Union (ITU) released the IMT-2030 framework [12], which established an official roadmap for 6G standardization. Notably, Integrated Sensing and Communication (ISAC) was introduced as a novel service scenario, becoming one of the six major application domains. ISAC enables communication signals to simultaneously provide information about the physical environment, a capability particularly valuable for cooperative communication and sensing (C&S) in UAV networks [13] and joint vehicle tracking and communication [14][15].

Full integration of communication and sensing functionalities requires coordinated optimization of hardware, signal processing pipelines, and waveforms, alongside efficient allocation of limited resources between these dual tasks [16][17]. The deployment of ISAC in UAV systems has garnered significant attention [18], as many advantages of UAV-enabled communications extend to sensing-related missions. UAV mobility enables repositioning for optimal sensing geometry, while high-probability line-of-sight (LoS) paths in UAV systems mitigate occlusion challenges in sensing domains. Additional discussions on ISAC in UAV contexts appear in works by Meng et al. [19] and Mu et al. [20]. Meanwhile, Li et al. [21] investigated optimal trajectory design for UAVs performing joint sensing and communication tasks with mobile ground nodes. However, existing research primarily focuses on UAVs being considered a promising aerial ISAC platform. Their high maneuverability and the potential to establish robust LoS links can enhance service coverage and performance [22][23][24]. Conversely, there is a dearth of literature specifically addressing UAVs acting as both sensing targets and communication terminals.

3) **Reconfigurable Intelligent Surface (RIS)**

The reconfigurable intelligent surface (RIS) consists of passive reflective elements, each with a configurable reflection/refraction coefficient. This capability allows for intelligent regulation of the electromagnetic (EM) waves, facilitating the reconfiguration of wireless channels. By coherently integrating desired signals and eliminating interference, communication throughput can be significantly enhanced without the need for additional active base stations (BS) or relays. It is also worth noting that RIS offers practical advantages, such as its lightweight nature, making it convenient to deploy on walls or fast-moving vehicles, thereby enabling a wide range of applications [25]. As a revolutionary technology, RIS effectively transforms the radio environment into an intelligent one, bringing substantial benefits to various sectors such as transportation, manufacturing, and smart cities. The usage of RIS has garnered attention as a promising technology for the development of 6G networks. RIS, characterized by its low cost, low power consumption, and low complexity, is a promising technology for achieving cost-effective UAV network coverage.

*B. Relevant Surveys and Our Contributions*

Recent years have witnessed a growing body of valuable surveys and tutorials on UAV communication research [26]-[33], as listed in Table I. In this context, this article takes a unique perspective by focusing on RIS-Enabled UAV Communications and Sensing, particularly addressing interference challenges in UAV networks arising from free-space path loss and atmospheric ducting effects.

TABLE I
RELEVANT SURVEYS AND MAGAZINES ON UAV AND RIS

| References | Focus |
|---|---|
| [26] | UAV serves as a communication UE or base station. |
| [27] | Optimization problems in UAV operation planning. |
| [28] | UAV serves as a BS. |
| [29] | Various 5G techniques based on UAV platforms. |
| [30] | Path planning for UAV communication networks. |
| [31] | UAV communication in 5G Networks. |
| [32] | Safety, traffic control, and security aspects for UAVs. |
| [33] | UAV regulations. |
| This review | This article takes a unique perspective by focusing on RIS-Enabled UAV Communications and Sensing, particularly addressing interference challenges in UAV networks arising from free-space path loss and atmospheric ducting effects. |

This paper focus on the second category, specifically addressing the challenges and solutions associated with achieving low-altitude coverage for UAVs using terrestrial networks [34]. The main contributions of this paper are as follows:

1) **Thorough Analysis of UAV Signal Coverage Challenges**

We provide a comprehensive examination of the key challenges associated with achieving low-altitude coverage for UAVs using terrestrial networks.

● **Dual-Layer Heterogeneous Network Topology:**
We introduce a dual-layer heterogeneous network topology integrating the terrestrial network coverage layer with the UAV low-altitude coverage layer, enabling 3D continuous coverage. This multi-layer design mitigates inter-layer interference while presenting unique challenges for network coexistence.

● **Channel Characteristics and Challenges in UAV-ISAC Networks:** We identify two primary channel characteristics that distinguish low-altitude UAV networks from terrestrial networks: LoS channels and multi-path fading. These unique channel properties introduce challenges related to sensing accuracy and interference management.

● **Challenges of Traditional Active Phased Array Antenna (APAA)-Based Networks:** Our analysis highlights the limitations of traditional phased array antenna-based networks for UAV signal coverage, focusing on issues of cost, complexity, and site selection. We propose two alternative approaches to address these limitations: enhancing existing BSs or deploying dedicated BSs for UAV coverage.

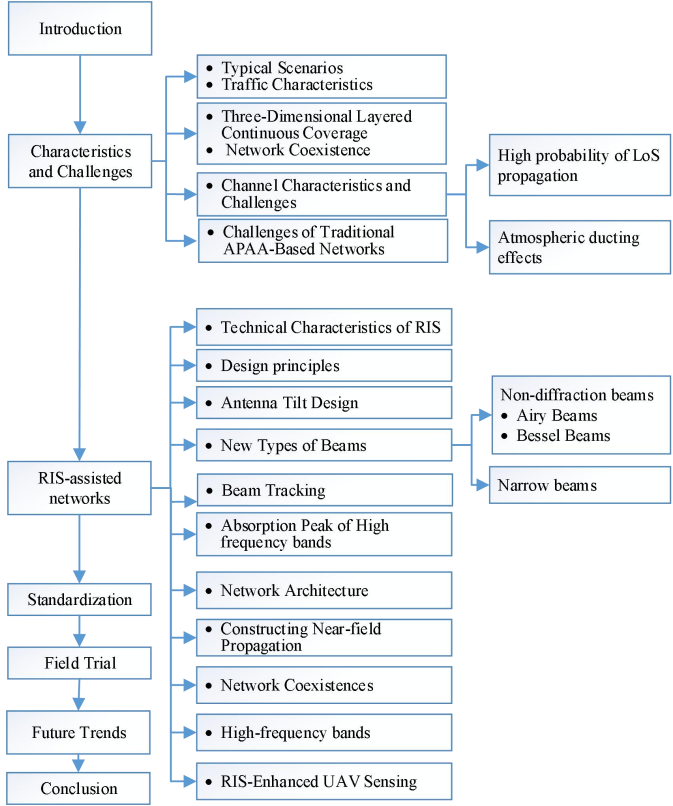

**Fig. 1.** Illustrates the organizational structure of the survey.

**2) RIS-Based Solutions for UAV Signal Coverage**

We propose RIS as a key solution to the challenges of UAV signal coverage. RIS enhances UAV-ISAC networks by improving signal propagation and providing continuous coverage.

● **RIS-Enabled UAV-ISAC Network Architectures:** We explore various RIS-based UAV network architectures, including cell-free designs and networks utilizing non-diffracting beams, aimed at increasing spatial degrees of freedom, managing interference, and enhancing sensing accuracy.

● **Millimeter-Wave (mmWave) Communications for UAV-ISAC Networks:** We discuss the potential of mmWave technologies for UAV communications and sensing, reviewing typical findings and challenges while exploring RIS-based solutions to address these issues.

**3) Standardization Aspects**

We address the standardization considerations for UAV-ISAC networks, emphasizing the need for scenarios and solutions that facilitate the integration of RIS with UAV communications.

**4) Field Trial and Future Technology Trends**

We present a field trial conducted in China to demonstrate RIS-based UAV coverage and outlines future research directions, including the variability of atmospheric ducting effects, emerging challenges in low-altitude UAV coverage, engineering implementation.

The structure of this paper is as follows (see **Fig. 1**). In Section II, we elaborate on the technical characteristics and challenges of UAV-ISAC networks. Building on the analysis in Section II, Section III proposes RIS-based solutions to address UAV signal coverage. Section IV explores the standardization aspects of UAV-ISAC. In Section V, we evaluate the performance of proposed solutions through field trials. Section VI outlines potential directions for future research. Finally, conclusions are drawn in Section VII.

## II. CHARACTERISTICS AND CHALLENGES OF UAV-ISAC NETWORKS

Although terrestrial wireless networks have been extensively studied for many years [35], their technological frameworks and empirical findings may not be directly applicable to future three-dimensional (3D) heterogeneous networks. These networks integrate both UAVs acting as aerial users/aerial-BSs and ground users, due to fundamentally distinct operational environments. Consequently, the new degrees of freedom introduced by UAV deployments remain largely unexplored, presenting unique challenges while simultaneously offering the transformative potential to enhance both communication and sensing capabilities.

### A. Typical Scenarios and Traffic Characteristics of ISAC Networks for UAVs

Compared to traditional terrestrial communication, UAV low-altitude coverage networks exhibit several distinctive characteristics. The unique features of ISAC networks in UAV-based scenarios are outlined as follows:

5) UAVs exhibit significantly higher mobility and greater freedom in trajectory flexibility compared to terrestrial nodes. This dynamic nature poses substantial challenges for channel estimation, beam tracking, beam switching, and cell handover processes.

6) While capacity requirements for UAV acting as UE and

sensing targets are generally lower than those for terrestrial coverage, the service distribution exhibits significant spatial-temporal variations.

7) ISAC networks for UAV applications represent typical ISAC scenarios, requiring simultaneous fulfillment of communication, sensing, and control service requirements. The sensing/positioning accuracy must surpass that of GPS/Beidou-based systems to justify the utilization of new intermediate frequencies and mmWave bands. Furthermore, for precise target detection, sensing operations should concentrate on the main lobe of the echoed signal while suppressing side lobes, thereby prioritizing the main lobe coverage for sensing services.

In contrast to traditional terrestrial cellular networks, ISAC networks designed for UAVs are characterized by their specific application scenarios, introducing entirely novel challenges that conventional cellular systems have not encountered.

### B. Dual-Layer Heterogeneous Network Topology for UAVs: Three-Dimensional Continuous Coverage and Network Coexistence

This section analyzes the fundamental differences in network coverage between terrestrial networks and low-altitude networks, with particular emphasis on the challenges of joint coverage optimization and the coexistence of multi-layer coverage networks.

The UAV dual-layer heterogeneous network topology integrates the terrestrial and UAV low-altitude network coverage layers to achieve three-dimensional continuous coverage [36]. The terrestrial network coverage layer relies on traditional cellular networks with appropriately configured down-tilted antennas to optimize signal reception ground users. In contrast, the UAV low-altitude coverage layer employs up-tilted antennas to enhance LoS transmission paths and ensure adequate coverage in low-altitude. This multi-layer design significantly reduces the coupling between the two layers, thereby mitigating inter-layer interference, and improving overall network performance.

Despite these advantages, the joint optimization of terrestrial and low-altitude coverage layers faces challenges.

1) **Cost-effectiveness:** Traditional cellular networks antennas with down-tilted are ill-suited for low-altitude UAVs. This increases hardware costs and power consumption due to the added system complexity. A balance must be struck between revenue from the sensing-related services and the network construction/operation expenses.

2) **Deployment complexity:** Dual antenna systems not only add the weight and size to the equipment, imposing higher payload requirements on BS systems, but also introduces challenges in site selection. The optimization must address BS layout, redundant construction minimizing, and cost reduction while maintaining effective coverage.

Beyond the coupling between terrestrial and low-altitude coverage layers, network coexistence between low-altitude UAV coverage layers and mid-to-high altitude coverage layers (e.g., civil aviation and low earth orbit satellites) requires attention [37]. UAV low-altitude coverage layers rely on LoS transmission with slow signal attenuation over distance, potentially causing spectrum competition/interference with mid-to-high altitude aircraft and other networks. Joint strategies must incorporate network coexistence factors to ensure harmonized layers interaction.

In summary, UAV low-altitude coverage research must comprehensively address cost-effectiveness, deployment complexity, and network coexistence.

### C. Channel Characteristics and Challenges of UAV-ISAC Networks

Compared to terrestrial coverage, ISAC networks for low-altitude UAV exhibit two primary channel characteristics: a high probability of LoS propagation and significant atmospheric ducting effects. Both characteristics result in low path loss, beneficial for wireless signal transmission. However, they may also lead to increased inter-cell and inter-network interference due to uncontrolled signal propagation range, diverging from the design principles of cellular networks. A fundamental tenet of the cellular network design is controlling coverage areas between neighboring BSs/sectors to enable small reuse factor and improve spectral efficiency.

1) **High probability of LoS propagation**

When UAVs operate airborne, they are typically unobstructed by ground obstacles (e.g., buildings, trees), resulting in predominantly LoS channels. The LoS-dominated air-ground channels inherent to low-altitude UAVs yield reduced path loss compared to terrestrial channels, owing to mitigated shadowing and multi-path fading [38]. Reference [39] presents a study that considers 3D mobility and LoS channel properties for cellular-connected UAVs. The authors identified key challenges and introduced analytical models with performance metrics based on quantitative evaluations.

In such scenarios, terrestrial obstacle shadowing becomes negligible, and the path loss closely approximates the free space propagation model which can be described by:

$$L_{fs}(dB) = 20\log_{10}(d) + 20\log_{10}(f) + 20\log_{10}(\frac{4\pi}{c}) \quad (1)$$

where $d$ is the signal propagation distance, $f$ is the signal frequency, and $c$ is the speed of light in vacuum.

This formula indicates that the primary factors affecting free-space propagation are the propagation distance and the carrier frequency. As the propagation distance increases, the path loss also increases. Similarly, higher signal frequencies result in greater path loss due to their shorter wavelengths, making high-frequency signals more susceptible to blockage by small objects during near-ground propagation. However, this effect is minimal in free-space propagation. **Fig. 2** compares the free-space path loss (including atmospheric and rain-induced attenuation) with the terrestrial urban path loss [40][41] under the following conditions: unit transmitter power, 28GHz mmWave operation, 64 transmit antennas, 16 receive antennas, optimal beamforming at both transmitter and receiver. Observations reveal 20~40dB lower attenuation for LoS channels compared to urban ground channels, with this divergence widening with distance.

In terrestrial wireless systems, buildings, vegetations, and other obstacles cause EM wave reflection, refraction, diffraction, scattering, limiting propagation range. This localizes cell coverage and confines signal to target areas (cells/sectors). In contrast, free space propagation enables longer distances and broader coverage due to the absence of terrestrial impediments. LoS-dominated environments further benefit sensing services. However, even with atmospheric and rain attenuation, quasi-free-space propagation remains nearly uncontrollable, posing challenges for UAV communications. Key issues include elevated co-channel interference between cells, reduced SINR, and frequent handover ping-pong effects. Moreover, far-field LoS-dominant MIMO channels exhibit rank-1 characteristics, hindering spatial multiplexing gains.

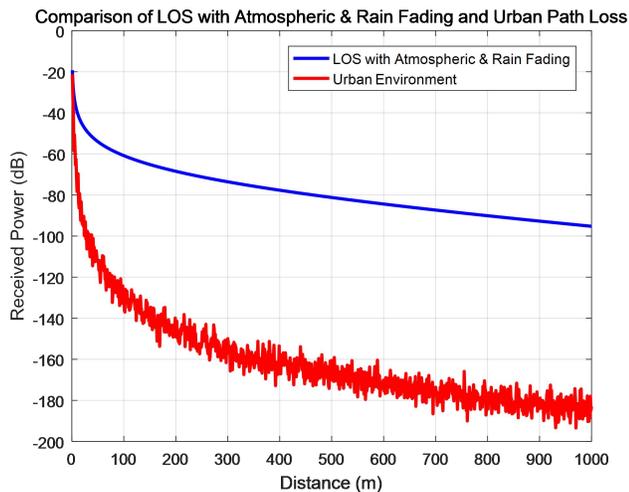

**Fig. 2.** Path loss comparison between LoS and Urban channel.

### 2) **Atmospheric ducting effects leading to long-distance propagation**

Atmospheric ducting emerges when vertical gradients in the atmospheric refractive index — primarily caused by temperature inversions or abrupt humidity variations — bend EM wave trajectories below the critical angle for total internal reflection. This phenomenon confines EM waves within a stratified atmospheric layer, where near-horizontal incidence angles induce repeated reflections between upper and lower boundaries. The resulting waveguide-like propagation traps wave energy through cyclical reflections, dramatically suppressing free-space spreading loss. Consequently, path loss in ducting environments becomes significantly lower than in normal atmospheric conditions, enabling anomalous long-distance signal transmission [42][43]. **Fig. 3** shows the atmospheric ducting phenomena.

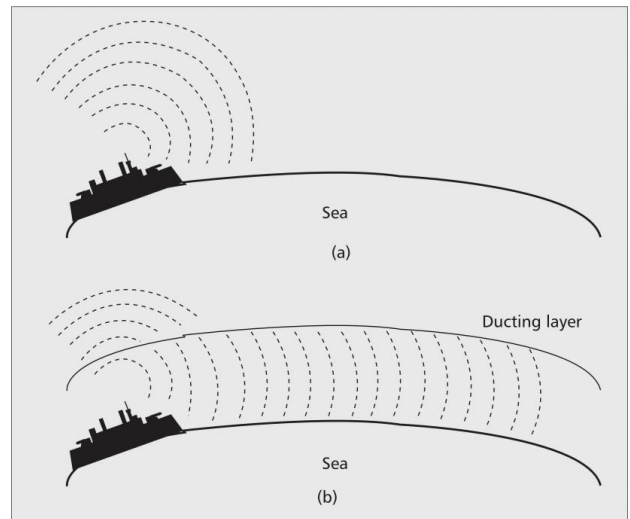

**Fig. 3.** Signal spreading in a) the standard atmosphere; b) the atmospheric duct [42].

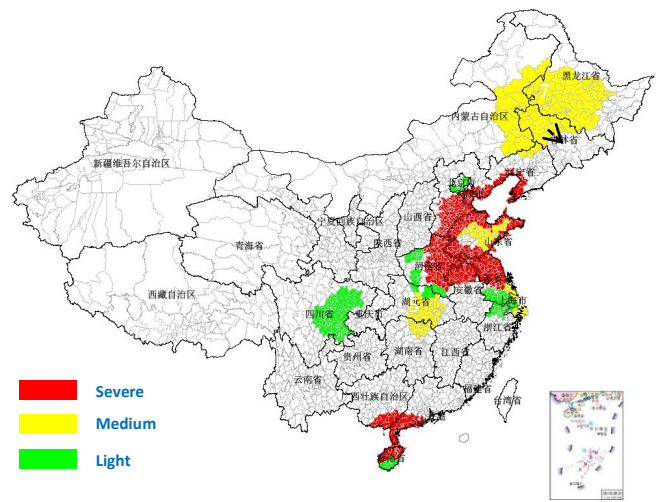

**Fig. 4.** Deployment of tropospheric ducting in China (TD-LTE networks).

**Fig. 4** illustrates the impact of atmospheric ducting phenomena observed in China Mobile's TD-LTE networks. The figure reveals tropospheric ducting occurrences in the Jianghan Plain, coastal regions of Hainan Island and Bohai Bay, and particularly in the North, Central, East, and Northeast China Plains. The website[2] provides 7-day global atmospheric duct forecast at 3-hour intervals for the initial 36 hours and 6-hour intervals thereafter. In wireless systems affected by atmospheric ducting, the interference range between victim and aggressor nodes spans 64-400km, typically remaining below 150km in most inland areas but occasionally exceeding 300km in coastal regions [43].

This phenomenon enables propagation distances far exceeding free-space propagation model predictions, thereby expanding UAV signal coverage while exacerbating inter-cell interference [44].

Parabolic equation (PE) methods derived from Maxwell's equations are widely employed to model atmospheric ducting channels [45]. By neglecting backscattering components, this approach reduces the three-dimensional wave equation to a computationally tractable two-dimensional forward propagation problem, enabling efficient modeling of complex atmospheric refractivity profiles. The channel path loss can be calculated using PE-based approaches through:

$$PL = 20\lg(\frac{4\pi}{|u(x,z)|}) + \lg(d) - 30\lg(\lambda) \qquad (2)$$

where $u(x,z)$ is the reduced field component, $x$ and $z$ are respectively the horizontal and vertical axes representing range and altitude, $d$ is the range from the transmitter and $\lambda$ is the wavelength [46].

Based on atmospheric ducting formation conditions, traditional terrestrial macro-BSs with small antenna down-tilt angles may inadvertently satisfy ducting criteria. In contrast, low-altitude UAV networks employing up-tilted antennas are inherently more susceptible to atmospheric ducting effects. Consequently, the interference impact of atmospheric ducting is more pronounced in UAV-oriented low-altitude coverage networks.

Similar to free-space fading effects, while atmospheric ducting extends the transmission range of wireless signals (benefiting scenarios like maritime long-distance coverage), it introduces challenges for conventional Time Division Duplex (TDD) systems [44]. Specifically, ducting-induced ultra-long-range propagation causes:

● Uplink and` downlink cross-link interference (false detection) between extremely distant BSs;

● Multi-path ghost sensing peaks (false alarms) from remote cell signals;

● The above analysis reveals distinct interference mechanisms between the two channel characteristics;

● LoS quasi-free-space fading primarily induces adjacent-cell co-channel interference;

● Atmospheric ducting enables ultra-long-distance signal propagation, causing inter-link interference in TDD systems.

### D. Challenges of Traditional APAA-Based Networks for UAV (Cost, Complexity, and Site Selection Challenges)

To extend coverage to low-altitude UAVs while maintaining traditional ground network coverage, it is necessary to deploy additional BSs equipped with up-ward angled phased array antennas to ensure continuous coverage for low-altitude UAVs, as shown in **Fig. 5**.

Two implementation approaches are viable:

**Approach 1:** Upgrading existing BSs

This method involves retrofitting existing BSs are upgraded with up-ward facing additional active phased array antennas (APAAs) for low-altitude coverage. While reducing the number of new BSs deployments and minimizing additional site acquisition requirements, this approach incurs substantial costs due to the APAA hardware. Furthermore, integrating these antennas into existing infrastructure introduces additional complexity, such as increased weight, volume, and power consumption demands, which may challenge the existing infrastructure.

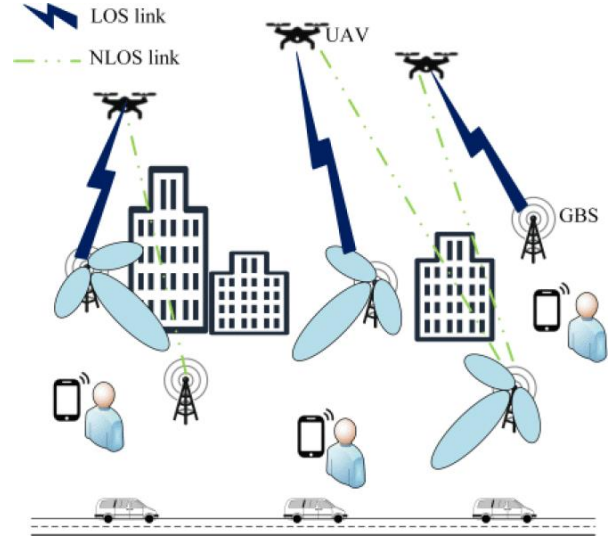

**Fig. 5.** The three-dimensional coverage situation of ground base stations (GBS) [47].

**Approach 2:** Dedicated BSs for UAV coverage

Alternatively, dedicated BSs specifically engineered for low-altitude UAV coverage may be deployed independently of existing infrastructure. These dedicated BSs incorporate upward-facing APAAs to achieve targeted aerial coverage. While this method enables optimized deployment unconstrained by existing BS site limitations, it necessitates substantial capital expenditure and presents significant site selection challenges.

Analysis of the two typical implementation approaches for APAA coverage using traditional BSs reveals that while both methods offer distinct advantages, they also face unique challenges related to cost, complexity, and site selection.

In summary, the characteristics and challenges of UAV networks are reflected in the following aspects:

1) Compared to ground coverage, low-altitude coverage exhibits high mobility and different traffic distribution, making it a typical application scenario for ISAC.

2) UAV networks represent a dual-layer heterogeneous network topology that requires three-dimensional, continuous coverage, posing challenges for multi-layer network coexistence.

3) Challenges arising from UAV channel characteristics include: LoS channels (free-space fading plus atmospheric and rain-induced fading), difficulty in controlling cell coverage areas, and potential increased co-channel interference between adjacent cells. UAV low-altitude coverage antennas require an up-tilted configuration, which, in contrast to traditional down-tilted, may lead to more pronounced atmospheric ducting phenomena. This can result in cross-link interference between extremely distant cells in TDD systems.

4) Challenges associated with traditional APAA-based BSs encompass cost, complexity, and site selection issues.

## III. RIS-ASSISTED NETWORKS FOR UAV SIGNAL COVERAGE

Due to their unique technical advantages, RIS are strong candidates for achieving cost-effective signal coverage for UAVs. Extensive research has been conducted on RIS in various system configurations [48][49]. These devices can be installed on the ground to facilitate UAV communications or mounted on UAVs to support ground communications [50]. This section explores innovative RIS-based approaches including the following: 1) antenna tilt optimization, 2) beam design, 3) utilization of high-frequency mmWave/terahertz frequencies, and 4) network architecture optimization. Each approach offers distinct advantages tailored to specific operational scenarios. Detailed examination of their theoretical foundations, design principles, and implementation considerations are provided.

### A. Technical Characteristics of RIS

RISs are best known for their low cost, low power consumption, low complexity, and ease of deployment. These advantages enable RISs to demonstrate extensive application potential in areas such as wireless channel enhancement, RIS-based phased array antennas, and information modulation [51]. Furthermore, with its ultra-large antenna aperture, RISs can readily create a ubiquitous near-field propagation environment for future 6G networks [52][53].

1) **Role of RIS in Future Networks: Auxiliary vs. Fundamental Network Element**

With the continuous evolution of network technologies, each generation introduces unique service characteristics and signature technologies. The 4G era is characterized by mobile broadband connectivity, with OFDM and MIMO as signature technologies, In the 5G era, the Internet of Everything and Massive MIMO emerged as signature technologies. For 6G, integrated sensing and communication, space-ground integration, and ubiquitous intelligence, RIS, new multiple access schemes, and other candidate technologies (e.g., AI-based solutions) hold potential to become the signature technologies.

**Auxiliary Network Element Role:** In specific scenarios, RIS acts as an auxiliary network element, enhancing performance and providing incremental improvements to existing networks.

**Fundamental Network Element Role:** In most cases, RIS serves as a fundamental network element, considering technology maturity, cost, power consumption, and complexity constraints. It provides essential capabilities for the practical deployment and large-scale commercialization of 6G networks — much like "offering timely assistance," — particularly in critical functionalities like virtual BSs.

2) **Typical Application Scenarios and Potential Opportunities of RIS in 5G-Advanced (5G-A) and 6G Networks**

a) **5G-A Scenarios**

Enhancement of low-frequency coverage: As an auxiliary network element, RIS can extend the coverage, effectively improving network performance.

Enabling continuous coverage for mmWave networks: In this scenario, RIS acts as a fundamental network element, ensuring continuous and reliable coverage for mmWave networks in power-limited environments.

b) **6G Network Scenarios**

Enabling continuous coverage in mid- to high-frequency and mmWave bands: RIS, as a fundamental network element, plays a critical role in ensuring seamless coverage of mid- to high-frequency and mmWave signals in 6G networks.

Enabling ubiquitous sensing and positioning: Through flexible RIS deployment, 6G networks can provide broader capabilities of sensing and positioning with high precision.

RIS-based phased array antennas: The integration of RIS technology enables innovative designs for phased array antennas in 6G networks, driving advancements in antenna technologies.

Enabling ubiquitous near-field communication: The deployment of RIS extends the coverage of 6G networks, especially in near-field communications.

In RIS-assisted networks, channel estimation and angle acquisition confront challenges from passive elements and high-dimensional cascaded channels. Key approaches include compressed sensing for sublinear pilot overhead via spatial sparsity, deep learning with hybrid networks to address near-field scattering, and hybrid active-passive architectures balancing cost and accuracy. Target angle detection employs beam-sweeping codebooks for static scenarios and subspace decomposition for multi-target resolution. System-level joint estimation frameworks optimize sparsity-constrained models via iterative algorithms, while scalability bottlenecks drive innovations in intelligent calibration and THz-band compensation.

In summary, the role of RIS technology in future networks will be flexibly adjusted based on the specific application demands. RIS can serve as either an auxiliary network element to enhance network performance or as a fundamental network element to support core radio network capabilities, thereby accelerating the evolution of 6G and beyond.

### B. RIS-Based UAV-ISAC Networks: Design principles

To effectively address the challenge of UAV signal coverage, the following design principles should be adhered to:

**Ensure Continuous Coverage:** Guarantee seamless signal coverage within the target area to avoid blind spots.

**Constrain Propagation Areas:** Limit signal propagation in non-target areas to reduce interference and enhance frequency reuse efficiency.

**Network Coexistence:** Implement a three-layer coverage network comprising the ground layer (e.g., traditional UEs and sensing targets), the low-altitude layer (e.g., UAVs), and the mid-to-high altitude layer (e.g., commercial aircraft).

**Reduce Cost and Complexity:** Consider deployment costs,

engineering complexity, and site selection constraints.

**Appropriate Access Capacity:** While RIS regulates channel propagation conditions to enhance throughput and sensing accuracy, it cannot increase terminal access or sensing capacity, which remain dependent on active BSs.

### C.  RIS-Based UAV-ISAC Networks: Antenna Tilt Design

Antenna tilt angle design is an effective technique for optimizing signal propagation, mitigating interference, and improving coverage by adjusting the antenna's vertical angle. In the context of UAV signal coverage, a carefully designed tilt angle can significantly enhance both signal range and quality, particularly in complex aerial propagation environments.

In current cellular networks, ground-based BS antennas are typically down-tilted, with their main lobes directed towards the ground to optimize coverage for terrestrial users [54]. This configuration adversely affects the UAV link, resulting in lower received signal power and reduced data rates at the UAV, as demonstrated by 3rd Generation Partnership Project (3GPP) studies [55]. UAVs flying above these BSs are predominantly served through the side lobes, posing technical challenges to providing reliable service to both UAVs and ground users simultaneously [56]-[59].

Consequently, the conventional down-tilted BS antennas—designed primarily for ground coverage and inter-cell interference suppression—must be reconfigured. 3GPP Studies confirm that UAVs receive weak signals from terrestrial BSs, necessitating further research and development to support aerial users effectively [60][61]. The authors of [62] highlighted the varying altitudes of UAV flight, while [62][63] addressed the importance of antenna tilting and 3D coverage models. A comprehensive study on the performance of various antenna deployments for cellular-connected UAV scenarios is presented in [64].

Two RIS-based deployment approaches for achieving low-altitude UAV coverage are discussed below [54][65].

#### 1)  Approach 1

BS antennas utilize conventional down-tilted configurations, while the non-collocated RIS regulates the electromagnetic wave propagation path and performs passive beamforming to direct signals towards low-altitude UAVs, as shown in **Fig. 6** [54]. In other words, the RIS optimizes channel conditions to facilitate upward signal transmission. By leveraging 3D passive beamforming via RIS reflections to serve in-flight UAVs, the architecture significantly enhances communication links between terrestrial nodes and UAVs, thereby minimizing modifications to the existing ground BS antenna infrastructure.

Regarding multiplicative fading in the RIS-cascaded channels, the primary path exhibits a LoS characteristics (as the RIS-UAV cascaded channel can easily establish LoS), which is predominantly modeled by free-space path loss. Consequently, signal attenuation increases more gradually with propagation distance, enabling effective signal coverage for low-altitude UAV, even under multiplicative fading conditions.

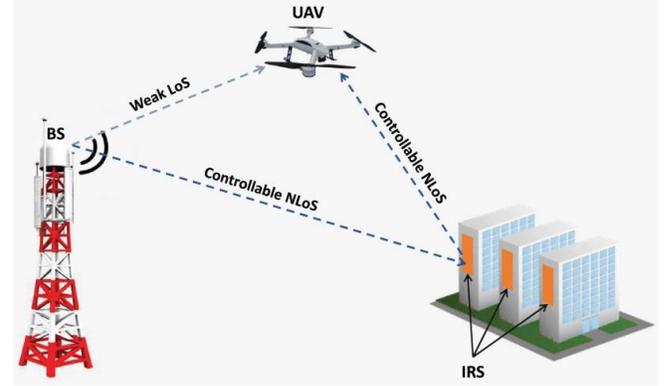

**Fig. 6.** System model for IRS-assisted cellular UAV communications [54].

The BS achieves simultaneous coverage for both terrestrial users and low-altitude UAVs using down-tilted antennas, thereby eliminating the necessity for additional up-tilted antenna installations. To mitigate upward propagation interference — specifically, neighboring cell interference caused by free-space fading and cross-link interference in TDD systems due to waveguide effects — the RIS-controlled beam should be angled as upward-tilting as possible. This minimizes horizontal interference leakage and avoids incidence angles that could induce total internal reflection conditions associated with waveguide propagation.

High-density deployment of RIS is required to suppress interference effectively. To achieve this, RIS beams should be oriented as narrowly as possible in the vertical plane, which inherently reduces the coverage area per RIS. Consequently, more RISs are required to maintain continuous UAV coverage. In [54], RIS has the potential to enhance cellular UAV communications, with received signal power gain analyzed as functions of UAV altitude and RIS parameters (including aperture size, deployment height, and BS separation distance). Using 3GPP ground-to-air channel models, the study reveals that a properly configured 100-element RIS can achieve 21 dB received power gain for a UAV hovering at 50 meters . Furthermore, optimal RIS deployment altitude decreases with increasing BS separation distance to ensure constructive signal reflection from the BS.

#### 2)  Approach 2

Building on Method 1, which utilizes relay-like RIS for UAV coverage, Method 2 introduces an additional upward-tilted antenna array at the BS to further meet UAV coverage. This configuration equips the BS with both down-tilted and upward-tilted antenna arrays. The upward-tilted array leverages RIS-based antennas, capitalizing on RIS's benefits (e.g., low cost, low power consumption, low profile, and low complexity) [66][67]. Similar to Method 1, the relay-like RIS deployed in the propagation environment controls the electromagnetic wave propagation path and beamforming for the down-tilted antennas, directing the enhanced signal to low-altitude UAVs. Method 2 incorporates two types of RIS deployments: (1) RIS-based phased array antennas mounted on the BS (i.e., co-located with the BS), and (2) relay-like RIS panels deployed in the propagation environment for channel regulation.

**Beam Upward-Tilt Constraint:** The tilt angles of the beams from both the BS and RIS [54] for UAVs should be as vertical as possible, thereby narrowing the coverage area of the BS and each RIS.

**Low-Power Transmission for Up-Tilted Antennas:** To mitigate interference and coexistence issues, the up-tilted antennas at the BS are configured for low-power transmission [68].

**Co-location of RIS-Based Up-Tilted Antenna Arrays:** The RIS-based up-tilted antenna arrays are mounted on the BS (i.e., co-located with the BS), overcoming challenges related to the weight, size, and power consumption of traditional APAA systems.

**High-Density RIS Deployment Requirement:** A sufficiently dense RIS deployment is necessary to ensure continuous UAV coverage. RIS's unique attributes (low cost, low power consumption, and low complexity) enable economically efficient, continuous UAV network coverage.

In summary, deploying RIS densely is crucial for mitigating interference and ensuring reliable UAV communication. While retrofitting RIS into existing sub-6 GHz 5G networks may be economically impractical due to installation costs, implementing RIS in new mid-band and mmWave networks is more feasible. This makes joint optimization of BS and RIS deployments a preferable approach.

### D. RIS-Based UAV-ISAC Networks: New Types of Beams

The primary objectives of beam design are to achieve efficient coverage of the target area by adjusting the direction and shape of the beam, while ensuring rapid signal attenuation in non-target areas to minimize interference. For UAV signal coverage, a well-designed beam can significantly improve both the range and quality of signal coverage, especially in the complex aerial propagation environments discussed in Section II.

The fundamental design concept of the new beams focuses on confining the coverage area to the intended target zone, thereby reducing interference with non-target regions.

**Beam Focusing:** Unlike far-field conditions, in near-field channel environments, beam focusing leverages the propagation characteristics of spherical waves to concentrate signal energy, enabling precise control of signal propagation in both angular and distance domains, as shown in **Fig. 7** [69][70]. This approach ensures concentration the beam within the target area to avoid excessive coverage and maintenance of adequate signal strength within the target area to meet communication requirements.

**Rapid Diffraction:** Outside the target area, the beam should quickly diffract, leading to signal strength reduction. By designing appropriate beam shapes (e.g., non-diffraction beams), the signal attenuate swiftly after leaving the target area, thereby minimizing interference with adjacent regions.

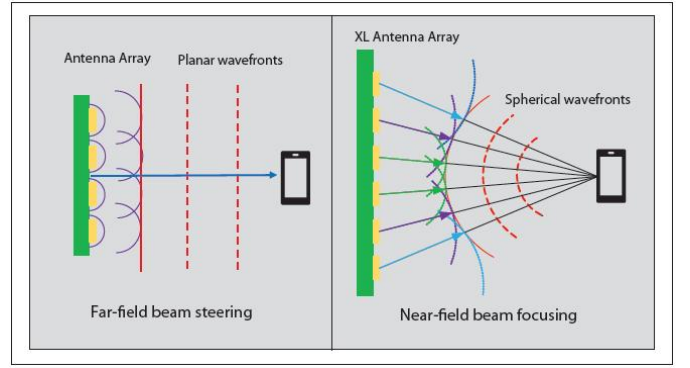

**Fig. 7.** IIlustration of the principles of far-field beam steering and near-field beam focusing [70].

**Narrow Beams:** Utilizing a super-large aperture RIS arrays to generate narrow beams reduces spatial overlap likelihood, effectively mitigating interference.

The utilization of RIS with ultra-large antenna apertures for novel beam pattern control offers effective interference mitigation in UAV networks. Diffraction-free beams (e.g., Bessel beams, Airy beams) and narrow beam characteristics represent innovative beam patterns for interference suppression in UAV networks.

#### 1) Non-diffraction beams

Diffraction-free beams [71], with their superior spatial characteristics, represent a novel approach for UAV coverage, and RIS with extremely large antenna apertures is a key candidate technology for realizing such beams.

**Bessel beams:** Bessel beams are a class of non-diffracting beams whose transverse electric field distribution is described by the Bessel function of the first kind. As exact solutions to the Helmholtz equation in cylindrical coordinates, Bessel beams exhibit a unique property as the central lobe maintains concentrated energy without diffraction spread during propagation, while the sidelobes decay oscillatory. This characteristic enables long-range stable coverage in UAV networks while suppressing multipath interference. The beams based on Bessel functions can concentrate signals in the target coverage area, reducing interference and leakage caused by signal diffraction to other angle areas.

Since their discovery in 1987, they have exhibited many interesting properties such as non-diffraction [76], self-reconstruction [77] and even providing optical pulling forces [78][79]. The scalar form of Bessel beams propagating along the $z$ axis can be described in cylindrical coordinates $(r, \varphi, \phi)$ by:

$$E(r, \varphi, \phi) = A \exp(ik_z z) J_n(k_r, r) \exp(\pm in\varphi) \qquad (3)$$

where $A$ is the amplitude, $k_z$ and $k_r$ are the corresponding longitudinal and transverse wavevectors. The Bessel function of the first kind $J_n(x)$ is a solution to the Bessel differential equation $x^2 \dfrac{d^2 y}{dx^2} + x \dfrac{dy}{dx} + (x^2 - n^2) y = 0$, where n is the order of the function. For zeroth-order Bessel beams, the transverse field distribution features a central main lobe surrounded by concentric annular sidelobes. The non-diffracting nature of Bessel beams arises from the mathematical property that the central lobe's energy density remains constant along the propagation axis, making them ideal for UAV communications requiring minimal signal leakage in non-target areas.

**Airy Beams:** Airy beams are asymmetric non-diffracting waves characterized by self-accelerating propagation and self-healing properties. Their transverse intensity profile follows the Airy function, which is the solution to the differential equation $\frac{d^2 y}{dx^2} - x \cdot y(x) = 0$. Unlike conventional beams, Airy beams exhibit parabolic trajectory bending during propagation while maintaining a non-diffracting core, making them ideal for dynamic coverage in UAV networks. The beams, known for its non-diffracting characteristics, can maintain high intensity within the target area while rapidly diffracting and scattering beyond it. Berry and Balazs [72] proposed that the Schrödinger equation of free particles can obtain the Airy beam envelope solution with infinite oscillation characteristics under the condition of paraxial approximation in 1979. This kind of beam, namely Airy beam, has remarkable characteristics of self-bending, non-diffraction, and self-healing, which subverts our traditional cognition of the light beam propagating along a straight line [73]-[75].

The electric field distribution of the Airy beams is given by the formula:

$$E(x,z) = Ai(\frac{x-z^2}{z_0^2})e^{i(\frac{xz}{2z_0^2} + \frac{z^3}{3z_0^3})} \qquad (4)$$

where $E(x,z)$ is the beam field strength, $Ai$ is the Airy function, $x$ and $z$ are spatial coordinates, and $z_0$ is a beam parameter. The Airy function is defined as the solution to the Airy differential equation. Its asymptotic behavior reveals exponential decay for x>0 and oscillatory decay for x<0. In the context of Airy beams, the main lobe corresponds to the first peak of Ai(x) in the x<0 region, while the sidelobes decay rapidly. This mathematical property enables two key physical phenomena including non-diffraction and self-acceleration with parabolic trajectory except external steering.

### a) Beam Direction Adjustment After Target Area

After the beam passes through the target area, the beam should propagate vertically, thereby reducing the horizontal transmission distance. Adjusting the beam's direction to propagate vertically after the target area helps prevent interference caused by horizontal transmission.

In UAV networks, the self-bending property allows Airy beams to dynamically adapt to UAV mobility. For instance, when a UAV ascends or descends, the parabolic trajectory of the Airy beam can automatically adjust the coverage area to follow the UAV's vertical movement, avoiding blind zones caused by fixed-beam misalignment. Additionally, the self-healing property ensures that partial obstructions do not permanently disrupt the beam profile, as the wavefront reconstructs itself after bypassing the obstacle — critical for maintaining reliable ISAC links in cluttered airspaces.

The directional gain of the beam is defined by the formula:

$$G(\theta) = G_0(\frac{\sin(N\pi\sin(\theta)/\lambda)}{N\sin(\pi\sin(\theta)/\lambda)})^2 \qquad (5)$$

where $G(\theta)$ is the gain in the direction $\theta$, $G_0$ is the maximum gain, $N$ is the number of antenna elements, and $\lambda$ is the wavelength.

### b) Composite Beams

Utilizing multiple beams in combination allows for more flexible control over signal coverage and interference. For example, two Airy beams with opposite phases can be combined to quickly diffract and diffuse energy after passing through a UAV. The electric field distribution for a multi-beam design is given by:

$$E_{total}(x,z) = \sum_n w_n E_n(x,z) \qquad (6)$$

where $E_{total}(x,z)$ is the total electric field distribution, $w_n$ is the weight of the $n$ th beam, and $E_n(x,z)$ is the electric field distribution of the $n$ th beam.

## 2) Narrow beams

In [80], UAV can fly much higher than BSs and buildings. Therefore, they tend to be visible to more cells. On the other hand, the communication between UAV and a BS may cause potential interference to other cells. Since the channels between UAV and BSs are in general dominated by LoS paths with small angle spread, the inter-cell interference can be effectively reduced by directional beamforming, which is also the enabling technique of mmWave communications. However, the beam-width needs to be carefully designed as narrow beams may lead to heavy training overhead for beam alignment, whereas wider beams may cause high interference.

## 3) General Steps for Beam Design in UAV Coverage

Designing an effective beam pattern is essential for optimizing UAV coverage. The general steps involved in this process are as follows:

a) **Determine Beam Parameters:** Define the beam-width, direction, and shape based on the coverage requirements of the target area. For applications requiring broader coverage, a wider beam may be selected, while for high-intensity coverage, a narrower beam is more suitable [81].

b) **Design Beam Shape:** Customize the beam shape according to the size and contours of the target area. For example, an Airy beam [73]-[75], known for its curved propagation characteristics, can address specific coverage needs, whereas a highly directional narrow beam is ideal for precise spatial coverage.

c) **Adjust Beam Direction:** Align the beam's propagation direction with the location of the target area. In aerial coverage scenarios, enhancing vertical propagation and minimizing horizontal components is often preferable [82].

d) **Optimize Beam Combination:** Use a combination of multiple beams to achieve flexible coverage. This

approach enables coverage of various target areas and helps control interference by integrating beams with different directions and shapes [82].

By carefully designing beam shapes and spatial propagation characteristics, signal coverage can be optimized, improving communication quality and overall system performance across diverse applications.

Beam design plays a pivotal role in optimizing UAV signal coverage. By fine-tuning the beam's direction and shape, effective coverage of the target area can be achieved, while minimizing signal propagation in non-target areas to reduce interference.

### E. RIS-Based UAV-ISAC Networks: Beam Tracking

In the context of ISAC scenarios for UAV networks, the design of beam tracking mechanisms holds paramount importance alongside the development of novel beamforming architectures. Relative to terrestrial networks, the high mobility of UAVs imposes stringent requirements on beam tracking. Channel estimation becomes computationally prohibitive in highly dynamic environments. In this context, accurate localization enables the utilization of geometric information for phase shift optimization and passive beam steering. Furthermore, precise positioning is critical for collision avoidance and safe navigation within dense urban canyons. Reference [83] introduced the simultaneous localization and phase shift (SLAPS) method, a millimeter-wave (mmWave)-based localization framework tailored for UAVs. For integrated sensing and communication (ISAC) in high-mobility scenarios, vehicle-to-everything (V2X) systems face challenges analogous to those encountered by UAV systems. The sensing-assisted beamforming paradigm for vehicle-to-infrastructure (V2I) communication is equally applicable to UAV scenarios. Reference [84] proposed two innovative schemes: a baseline approach where beamwidth is dynamically adjusted to fully encompass the vehicle, followed by an extended Kalman filter (EKF) for positional prediction and tracking of communication-relevant scatterers (CRS); and an enhanced approach that partitions each transmission block into two sequential stages.

The considered UAV network is inherently dynamic, posing challenges to traditional optimization methods. To address these issues, researchers in [85], [86], and [87] adopted deep reinforcement learning (DRL) algorithms such as Double Deep Q-Learning (DDQN) and Deep Deterministic Policy Gradient (DDPG) for real-time optimization of UAV trajectories and passive beamforming. UAVs enhance both communication (by providing superior data transmission links) and sensing capabilities (through refined signal control). This necessitates a robust design optimization framework for RIS-empowered UAV-assisted ISAC systems using a novel RL approach [88] — transfer learning-enhanced DDPG — which accommodates multiple non-stationary targets, multiple users, and imperfect channel state information (CSI).

In RIS-enabled UAV-ISAC networks operating within mid- and high-frequency bands, near-field propagation emerges as a characteristic operational scenario [89]. Conventional far-field beam training methodologies typically focus solely on angular-domain beam optimization. However, near-field beam training necessitates joint optimization across both angular and range domains, rendering traditional exhaustive beam search approaches computationally prohibitive due to excessive latency overhead [93]. This necessitates the development of low-overhead near-field beam training protocols. To mitigate pilot contamination and training overhead in near-field regimes, deep learning (DL)-based beam training frameworks have been proposed in [94] and [95], leveraging spatial-temporal correlation models for dimensionality reduction. For cascaded RIS channel environments characterized by multi-hop signal propagation, reference [96] introduces a novel cascaded channel decoupling strategy that disentangles direct and reflected path components through tensor decomposition, thereby addressing the inherent parameter coupling challenges in RIS-assisted systems.

### F. RIS-Based UAV-ISAC Networks: Absorption Peaks of High frequency bands

Specific absorption peaks in high-frequency bands exhibit significant atmospheric absorption, leading to rapid signal attenuation with distance. Leveraging this attenuation characteristic enables effective confinement of the signal coverage area.

In modern wireless communication systems, where spectrum resources are scarce, interference management is critical. The mmWave and terahertz (THz) bands, with their extensive spectrum bandwidth, are considered key technologies for meeting future communication demands. Specifically, the mmWave and THz bands in the 30 GHz to 300 GHz range have garnered attention due to their unique propagation characteristics. Certain absorption peaks within these bands, such as those caused by atmospheric oxygen and water vapor, result in rapid signal attenuation with distance. This phenomenon provides a novel approach to managing coverage and interference in target areas.

#### 1) Physical Properties of Absorption Peak Bands

The absorption peak bands exhibit significant absorption characteristics in the atmosphere. Molecules like oxygen and water vapor in the atmosphere exhibit strong absorption in these bands, leading to rapid signal attenuation with propagation distance. For example, frequencies around 60 GHz and 183 GHz are typical absorption peak bands [97][98]. This absorption phenomenon can be represented by the following formula:

$$P(d) = P_0 e^{-\alpha d} \tag{7}$$

where $P(d)$ is the signal power at a distance $d$, $P_0$ is the initial transmitted power, and $\alpha$ is the atmospheric absorption coefficient, which depends on frequency and atmospheric conditions.

To further explore the physical properties of the absorption peak bands, the following aspects can be considered:

**Absorption Mechanism:** In the high-frequency mmWave and THz bands, electromagnetic wave interaction with atmospheric molecules (e.g., oxygen and water vapor) causes significant absorption. This arises from transitions in vibrational and rotational energy levels, which absorb electromagnetic waves energy, resulting in signal attenuation. For instance, the 60 GHz and 183 GHz bands correspond to absorption peaks for oxygen and water vapor, respectively.

**Absorption Coefficient:** The absorption coefficient ($\alpha$) is a key parameter describing the absorption characteristics of electromagnetic waves in the atmosphere. It depends on frequency, meteorological conditions (e.g., temperature, humidity), and atmospheric composition. Experimental measurements and theoretical models determine $\alpha$. Typical values are $\alpha \approx 0.15 km^{-1}$ at 60 GHz and $\alpha \approx 1.5 km^{-1}$ at 183 GHz bands, reflecting frequency-dependent absorption intensity.

2) **Constraining Coverage Area**

Utilizing absorption peak band characteristics, the coverage area can be precisely constrained by adjusting transmission power and selecting appropriate frequencies, enabling optimized communication system design.

To effectively constrain the signal coverage area, optimization can be approached from two perspectives:

a) **Adjusting Transmission Power:** Dynamically controlling power allows flexible coverage radius adjustment based on target area size and required range.

b) **Frequency Selection:** Higher frequencies (e.g., 183 GHz) with higher $\alpha$ values yield smaller coverage radius, suitable for precise small-areas coverage. Conversely, lower frequencies, for example (e.g, 60 GHz) with lower $\alpha$ values provide larger coverage areas, ideal for extensive regions.

*G. RIS-Based UAV-ISAC Networks: Network Architectures*

Network architecture plays a critical role in improving the coverage and quality of UAV signals. By optimizing cell layouts, deploying distributed antenna systems, and incorporating RIS, networks can achieve efficient signal coverage and reduce inter-cell interference.

As discussed in Section II, UAV network coverage adopts a dual-layer heterogeneous topology, combining the ground network coverage with the low-altitude UAV network coverage to provide continuous 3D service. Notably, the traffic distribution in low-altitude UAV coverage differs significantly from that of ground coverage. Optimizing the RIS-based network architecture is essential to accommodate the heterogeneous characteristics of the dual-layer topology and its unique traffic distribution.

In contrast to terrestrial coverage scenarios, low-altitude UAV terminals exhibit relatively sparse spatial density, resulting in lower per-unit-area capacity/access requirements and fewer sensing targets. The service profiles of UAV communications also diverge significantly from terrestrial counterparts: while terrestrial terminals support heterogeneous traffic types (e.g., voice, data, multimedia), UAV communication primarily involves control signaling or video streaming with relatively uniform traffic patterns. From a sensing perspective, terrestrial targets demonstrate greater diversity in electromagnetic characteristics and mobility constraints, whereas UAVs present as more homogeneous targets with predictable kinematic behaviors. Spatial distribution analysis reveals marked differences: terrestrial mobile targets operate within confined environments subject to physical infrastructure limitations, while low-altitude UAVs enjoy three-dimensional maneuverability with flexible trajectory adaptation capabilities. Even in planned flight path scenarios, UAVs retain greater flexibility in maneuvering along their routes compared to terrestrial vehicles constrained by road networks. Furthermore, as previously noted, low-altitude UAV coverage employs a quasi-free-space propagation model characterized by minimal diffraction and scattering losses. This propagation environment induces strong inter-cell coupling effects in cellular networks. Without proper inter-cell interference coordination (ICIC) mechanisms, adjacent cells may experience severe co-channel interference and excessive handover events, particularly in high-mobility UAV scenarios where Doppler shifts and channel aging effects exacerbate the problem [54].

To overcome the issues related to inter-cell interference and handover in traditional cellular networks, the so-called cell-free (CF) massive MIMO network infrastructure has been proposed as a promising approach for realizing the general distributed massive MIMO concept [99], where cell boundaries do not exist. Ubiquitous cell-free Massive MIMO refers to a distributed Massive MIMO system that implements coherent user-centric transmission to overcome the inter-cell interference limitation in cellular networks and provide additional macro-diversity. These features, combined with the system scalability inherent in the Massive MIMO design, distinguish ubiquitous cell-free Massive MIMO from existing coordinated distributed wireless systems [100]. Authors investigate the enormous potential of this technology while addressing practical deployment issues to deal with the increased back/front-hauling overhead deriving from the signal co-processing. In cell-free systems, a large number of geographically distributed antennas jointly serve a smaller number of UEs via a fronthaul network and a central processing unit (CPU), all of which operate on the same time-frequency resource. The cell-free massive MIMO concept builds on earlier distributed antenna systems (DAS) [101], network MIMO, and coordinated multipoint (CoMP) [102], all of which belong to the category of distributed antenna network architectures. Owing to its inherent advantages, cell-free massive MIMO is regarded as a key enabling technology for 6G networks.

Given its flexibility, the cell-free architecture is well-suited for UAV coverage in dual-layer heterogeneous networks. It enables flexible coordination access points (APs) to support a broader UAV coverage area while using fewer APs to cover smaller ground areas, aligning with heterogeneous network structure. Additionally, since the low-altitude UAV coverage relies on beam angles from each antenna arrays, multi-node cooperation is essential for continuous coverage of the target area, making the cell-free architecture a natural fit.

Since the advent of 4G LTE, CoMP technology (a predecessor to cell-free) has been standardized, though widespread commercialization remains limited. CoMP operates in two modes: (1) CoMP-JT, Multiple APs Joint Transmission, where various signal components of a user terminal (UE) are managed by different access points (APs) and traverse different propagation paths; and (2) CoMP-DPS, Dynamic Selection of a Single AP to Transmit a UE Signal. Furthermore, CoMP-JT can be further divided into coherent and non-coherent CoMP-JT. Coherent CoMP-JT refers to signals from multiple coordinated transmission points (e.g., base stations or antennas) that are synchronized in both phase and timing. Non-coherent CoMP-JT refers to signals that are transmitted independently, without phase alignment [104]. While CoMP was considered for 5G, current implementations only involve non-coherent CoMP JT. The primary advantage of CoMP lies in coherent CoMP JT, yet the benefits achieved through non-coherent CoMP JT are limited, posing challenges for broad commercial applications. One major obstacle to implementing coherent CoMP with traditional active phased arrays is the complex issue of antenna reciprocity calibration between different APs [103]. Similar to CoMP, cell-free systems face the same challenge of antenna reciprocity calibration when implementing coherent transmission. However, RISs, which utilize passive control, do not face the inherent challenges of antenna channel reciprocity calibration, providing a feasible solution for enabling coherent transmission in cell-free networks [104].

Furthermore, [104] addresses phase differences between RIS elements, proposing BS-coordinated by the can align the phase alignment across multiple RIS units to enable coherent transmission in cell-free systems. Integrating cell-free and RIS technologies can significantly enhance UAV communication coverage and quality, improving overall system performance.

### H. RIS-Based UAV-ISAC Networks: Constructing Near-field Propagation to Increase Spatial DoF

Traditional wireless communication networks primarily operate in sub-6 GHz spectrum bands, and even sub-3 GHz in some cases. Limited by low-dimensional antenna arrays and relatively low operating frequencies, the wireless near-field range is typically confined to within a few meters or even centimeters. Consequently, traditional wireless communication systems can be effectively designed under the far-field plane wave assumption. However, considering the large aperture of Extremely Large-Scale Antenna Arrays (ELAA) and the utilization of higher frequency bands in 6G networks, the near-field region can extend to tens or even hundreds of meters, rendering the conventional far-field plane wave assumption obsolete [105]. Therefore, in 6G networks, the near-field region becomes non-negligible, necessitating research into new near-field technology paradigms. From the perspective of spatial dimensional resource utilization, traditional wireless communication systems have already fully exploited far-field spatial resources. Further exploration and utilization of near-field spatial resources hold the potential to introduce new physical spatial dimensions for wireless communication systems. Reconfigurable Intelligent Surfaces (RIS), characterized by their large aperture, passive anomalous reflection capabilities, low cost, low power consumption, and ease of deployment, offer the opportunity to create a ubiquitous near-field wireless propagation environment in future 6G networks and give rise to entirely new network paradigms. In recent years, research on near-field propagation characteristics has gained significant traction, with rapid advancements being made. In particular, RIS-based near-field technologies have emerged as a research hotspot, yielding numerous notable achievements [106].

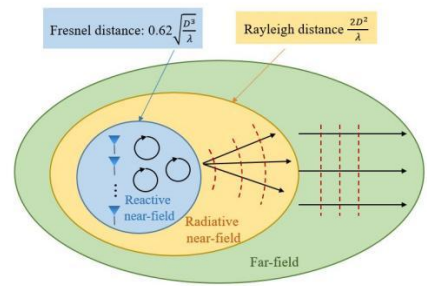

**Fig. 8 Boundaries of various propagation regions and characteristics of near-field and far-field wavefronts** [108].

As depicted in Fig. 8, according to electromagnetic field and antenna theories, the fields surrounding a transmitter can be categorized into near-field and far-field regions. The near-field region is further subdivided into the reactive near-field and radiative near-field. The reactive near-field region is situated in close proximity to the antenna, within the Fresnel distance, while the radiative near-field region extends a few wavelengths from the antenna, lying between the Fresnel and Rayleigh distances. The classical boundary between the near-field and far-field regions is defined by the Fraunhofer or Rayleigh distance [108], where the maximum phase difference between the spherical and plane wave models does not exceed $\pi/8$. This distance is calculated as $2D^2/\lambda$, with D representing the maximum aperture of the antenna and $\lambda$ denoting the carrier wavelength.

In the near-field region, within the radiative near-field, the amplitude of electromagnetic waves varies smoothly across different antennas in an array, whereas phase shifts increase significantly and exponentially with the antenna index [109]. Consequently, the signal propagation model in this region must be described using a spherical wave model. In contrast, the far-field region permits a simpler plane wave approximation. Traditional mechanisms based on far-field assumptions, such as channel estimation, beam training, codebook design, and beamforming, require redesign and optimization to align with near-field propagation characteristics. While far-field LoS conditions typically provide only one DoF, near-field spherical waves exhibit non-linear phase shifts and varying power levels across different links, leading to increased diversity and a higher rank of the channel matrix.

The LoS channel of UAVs is advantageous for sensing, albeit at the cost of reduced spatial multiplexing gain in MIMO systems. RIS, with its advantages of low cost, low power consumption, and flexible deployment, presents a potential solution for achieving continuous coverage in the new mid-frequency and mmWave bands. For example, reusing existing 5G BS sites (i.e., deploying mmWave BSs with a density similar to or slightly higher than that of existing sub-6 GHz BSs) and densely deploying RIS between BSs can enable cost-effective continuous coverage in these frequency bands. In the mid-to-high frequency bands, the combination of RIS's ultra-large aperture and dense deployment significantly enhances near-field characteristics, thereby increasing spatial degrees of freedom and improving spatial multiplexing gain [89][110].

The combination of the ultra-large aperture of RIS and its dense deployment in the mid-to-high frequency range will create ubiquitous near-field propagation, supporting broadband high-throughput communication and enabling continuous coverage for high-precision sensing [69].

*I.  RIS-Based UAV-ISAC Networks: Network Coexistence*

Reference [111] explores spectrum management strategies for UAV operations, identifying schemes that align with UAV characteristics and spectrum requirements, and assumes coexistence with existing wireless technologies occupying the spectrum. It also presents the rulings from policymakers and regulators and discusses the operational bands and radio interfaces.

Coexistence challenges between low-altitude and mid-to-high altitude coverage involve coordinating various flight altitudes, communication requirements, and coverage areas. To ensure efficient and interference-free coexistence of these networks, in-depth research is needed in areas such as spectrum management, interference mitigation, and network architecture optimization.

**Layered Coverage:** Given the significant differences in service demands between low-altitude and mid-to-high altitude coverage, adopting a layered network coverage approach is the most suitable choice.

**Network Coexistence:** Altitude differences and frequency overlaps between these layers may lead to mutual interference, necessitating the consideration of coexistence strategies among different network layers. Low-altitude coverage must address coexistence with ground networks and mid-to-high altitude networks (e.g., commercial aircraft). Furthermore, the introduction of RIS brings unique network coexistence challenges. Existing RIS typically exhibit broadband characteristics, which not only tune signals within the target spectrum but also affect non-target signals, leading to significant network coexistence issues [112][113]. References [112]-[116] propose various effective solutions to address these challenges.

The ubiquitous deployment of RIS enables dynamic electromagnetic wave control, effectively managing interference and facilitating the coexistence of UAV networks.

*J.  RIS-Based UAV-ISAC Networks: High-frequency Bands (mmWave)*

mmWave technologies for UAV have garnered significant attention, with numerous research findings published. This section reviews typical findings and discusses RIS-based communications and sensing for UAVs in mmWave bands.

The survey in [117] provides an overview of related works in UAV communications and technology integration. It explores mmWave beamforming-enabled UAV communications, addressing both technical potential and challenges, as well as relevant mmWave antenna structures and channel modelling. Additionally, technologies and solutions for UAV-connected mmWave cellular networks and mmWave-UAV ad hoc networks are reviewed.

UAVs operating in the mmWave spectrum represent an exciting frontier in wireless communication technology. However, UAV mmWave communication presents several significant challenges that need to be addressed for successful implementation and reliable operation [118][119].

Notably, in the future development of wireless networks, the introduction of RIS technology offers an innovative solution to the problem of high-frequency path loss. RIS, with its remarkable advantages of low cost, low power consumption, and low complexity, exhibits immense potential in fully utilizing the bandwidth of mmWave and THz bands. Through the application of RIS, we can unlock new possibilities in high-frequency spectral resources utilization, inaugurating a new paradigm in high-frequency resource utilization.

1) *Free Space Propagation Model—Friis Formula*

This model assumes that radio waves propagate in a vacuum, free from obstacles and interference, with received power decreasing exponentially with distance. This attenuation can be described by the Friis formula [120],

$$P_r/P_t = A_r A_t /(d^2 \lambda^2) \qquad (8)$$

where, $P_t$ is power fed into the transmitting antenna at its input terminals, $P_r$ is power available at the output terminals of the receiving antenna, $A_r$ is effective area of the receiving antenna, $A_t$ is effective area of the transmitting antenna, $d$ is distance between antennas, $\lambda$ is wavelength.

The Friis formula, also known as the Friis Free Space Formula or Friis Transmission Equation, is the fundamental equation describing the free space propagation model. Proposed by Danish-American radio engineer Harald T. Friis in 1946 [121], it closely links transmission power, antenna gain, distance, wavelength, and received power, making it one of the most important equations in antenna theory.

Friis formula shows that the relationship between frequency and free-space fading is based on the assumption of a simplified isotropic antenna effective area $A_{er}$, which is directly related to the type of antenna. However, the definition of path loss clearly states that "path loss is a measure of the reduction in signal strength due to distance and obstacles during the propagation of radio waves," meaning it should fundamentally be concerned only with the propagation of radio waves and not directly with the specific factors of the transmitting and receiving antennas. In other words, ideally, we consider only the effects of the propagation path of the electromagnetic wave, treating the transmitting and receiving antennas as independent of the influence of the propagation path. In other words, when considering electromagnetic wave propagation alone without accounting for the characteristics of the transmitting and receiving antennas, free-space path loss is independent of frequency. However, the free-space path loss model defined by the Friis formula explicitly considers the characteristics of the receiving antenna.

2) **RIS Paving the Way for a New Paradigm in High-Frequency Resource Utilization**

In practical engineering, the three core constraints in antenna design are cost, power consumption, and complexity, while also considering the physical space and load-bearing capacity of the deployment environment. As the frequency increases, a notable trend is that the ratio of the antenna's effective aperture to its physical size tends to decrease, meaning that to maintain the same effective aperture, the physical size of the antenna must be enlarged as the frequency increases. Additionally, due to current technological limitations, the manufacturing cost and complexity of individual antenna elements in high-frequency bands also increase. These two factors combined make it more challenging to achieve equivalent performance in high-frequency antennas, leading to higher costs and more complex system designs.

However, the growing attention to meta-surface technologies, particularly RIS, offers an innovative solution to this problem with its characteristics of low cost, low power consumption, low complexity, lightweight design, and ease of deployment. By introducing RIS, a new material and technology, novel phased array antennas based on intelligent meta-surfaces have emerged, offering a promising approach to achieving larger effective apertures.

In fact, prototype mmWave and sub-THz phased array antennas based on RIS have already been successfully developed and empirically validated, with their outstanding performance fully demonstrating the feasibility and potential of this technological pathway [66][67]. Looking ahead, in the deployment of 5G-A and 6G networks, and considering the multiple challenges of cost, power consumption, complexity, and deployment difficulty, RIS-based novel phased array antennas are poised to be game-changers by enabling frequency-independent effective apertures. This could effectively mitigate or even eliminate concerns about path loss in high-frequency bands (particularly mmWave and THz bands) under the Friis path loss model. This implies that future wireless networks will be able to fully exploit the abundant spectrum resources of high-frequency bands without being hampered by the long-standing prejudice of "excessive path loss," thereby ushering in a new paradigm of high-frequency resource utilization.

### K. RIS-Based UAV-ISAC Networks: RIS-Enhanced UAV Sensing

As elaborated in Section I, the deployment of UAVs necessitates enhanced sensing capabilities to enable precise control and regulatory oversight. The integration of RIS significantly improves UAV sensing performance through three primary mechanisms:

**LoS Channel Construction:** Leveraging the easy of deployment and configuration of RIS, stable LoS communication channels can be readily established [53]. This not only facilitates high-precision sensing but also simplifies algorithm design for LoS-dominant environments.

**Near-Field Spatial Sensing:** As discussed in Section III(H), when combined with high-frequency carriers, ultra-large aperture RIS arrays create near-field propagation environments [89]. These configurations exploit spatial resolution characteristics to enable millimeter-level UAV localization and trajectory tracking.

**Sensing-Integrated Architectures:** By incorporating embedded-sensing modules into RIS designs, joint communication and sensing (JCAS) frameworks can achieve real-time dynamic tracking of high-mobility UAVs, addressing challenges such as Doppler shift compensation and beam coherence in fast-varying channels.

Reference [90] proposes a joint optimization method for constant-modulus waveforms and discrete RIS phase shifts while satisfying the Cramér-Rao bound (CRB) constraint for target Direction-of-Arrival (DoA) estimation in RIS-assisted ISAC systems. The constant-modulus waveform enhances both radar and communication system performance, particularly in environments sensitive to multiple targets or interference [91]. This work jointly optimizes radar and RIS beamforming to maximize communication signal-to-interference-plus-noise ratio (SINR) while ensuring radar detection performance [92]. Additionally, it also co-optimizes active beamforming at base stations and passive beamforming at the RIS to maximize the total achievable throughput for communication users in ISAC systems. Deploying RIS in ISAC systems can significantly improve overall performance with marginal trade-offs compared to communication-only systems, although it requires balancing communication and radar detection performance requirements.

Furthermore, RIS-assisted UAV-ISAC networks represent a canonical multi-user and multi-target operational paradigm. Specifically, RIS is strategically employed to facilitate joint transmission from multiple BSs to multiple users, while concurrently enabling cooperative sensing across multiple BSs for multi-target detection. Reference [122] proposes an RIS-assisted cooperative multi-cell ISAC framework, where RIS enhances both joint transmission to the multiple users and joint active/ passive sensing for multiple targets.

Authors in [123] proposed a novel ISAC mechanism that accounts for the asymmetry between detection frequency and communication requirements. They jointly optimize UAV trajectories, user associations, target detection scheduling, and transmit beamforming to enable flexible trade-offs among detection frequency, power allocation for detection, and communication performance metrics — thereby maximizing the system's achievable throughput.

## IV. STANDARDIZATION

Standardization efforts are crucial for enabling large-scale commercial deployment of UAV communication technologies. Recent initiatives by standardization bodies have accelerated the development of unified frameworks for UAV connectivity.

Since the 4G LTE standard phase, UAV connectivity has been a focal research area within the 3GPP [124]-[126]. Notably, 3GPP Release 14 established explicit requirements for UAV connectivity, mandating continuous cellular network connection for aerial vehicles operating at velocities up to 300 km/h [124].

Building upon these foundations, 3GPP has made substantial contributions to 5G New Radio (NR) standardization to address UAV-specific connectivity demands [127][128]. In 2017, a dedicated study item was approved to enhance LTE support for UAVs, with particular focus on identifying technical challenges associated with maintaining connectivity using terrestrial networks equipped with down-tilted base station antennas. To further advance UAV support, 3GPP Release 17 introduced 5G network capabilities targeting connectivity enhancement, identification, and tracking requirements [128]. Several other standardization bodies and working groups have also dedicated substantial efforts to developing UAV-specific regulations [129], including:

- The International Telecommunication Union Telecommunication Standardization Sector (ITU-T) [130];
- The European Telecommunications Standards Institute (ETSI) [131];
- The IEEE Drone Working Group (DWG).

To accelerate research and innovation in aerial communications, the IEEE Vehicular Technology Society (VTS) established a Drone Ad Hoc Committee, while the IEEE Communications Society (ComSoc) launched an Emerging Technologies Initiative focused on aerial users and networks [132][133]. A key application area of this initiative is public safety, encompassing:

- Emergency cellular coverage extension via aerial platforms;
- Advanced service delivery for first responders during critical incidents.

Two joint working groups, initiated by IEEE ComSoc and IEEE VTS, are currently developing standards for aerial communication systems and networks [134]. In the 3GPP TSG RAN #80 meeting, a study item "Study on remote interference management for NR" was approved for 5G Release 16 regarding the atmospheric duct [135]. The work item "Cross Link Interference (CLI) handling and Remote Interference Management (RIM) for NR" is issued by 3GPP. In such work item, remote interference management (RIM) aims to investigate possible techniques for mitigating the impact of remote interference caused by atmospheric duct.

Additionally, 5G-A Rel-18 was completed in June 2024. Its ISAC standardization content servers as a foundation for UAV-ISAC standardization. The Rel-19 ISAC channel modeling can also serve as a basis for research on UAV-ISAC channel modeling [136], once its standardization work is completed.

In the future 6G standardization process, it will be necessary to address not only the individual standardization efforts for UAV communications and RIS, but also the specific standardization work that considers the integration of RIS and UAV communications. At the very least, scenarios and relevant solutions for RIS-enabled UAV-ISAC networks, as discussed in the previous sections, should be considered.

## V. RIS-BASED UAV COVERAGE FIELD TRIAL

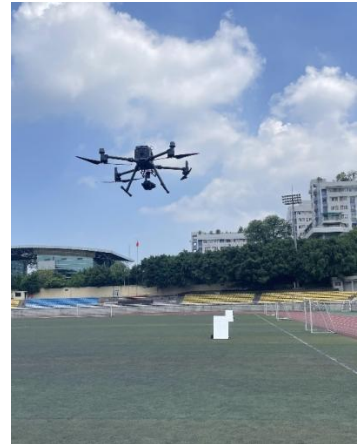

**Fig. 9.** RIS-based UAV coverage field trial configuration.

A field trial was recently conducted in Chongqing, China, to validate RIS-assisted UAV coverage performance. The test site comprised a university campus playground surrounded by undulating terrain. A quadcopter UAV was deployed at an altitude of 120 m, following a predefined grid trajectory covering the entire experimental area. The BS, operating in dual-band configuration (2.6 GHz and 4.9 GHz), was positioned 150 m from the playground center. Two RIS panels were evaluated during the trial:

- **2.6 GHz Panel:** 32×16 element array (512 total elements);
- **4.9 GHz Panel:** 32×32 element array (1024 total elements).

Both panels were strategically placed near the playground center, mounted horizontally (0° elevation angle) as illustrated in **Fig. 9**.

Despite operating the RIS in a semi-static configuration mode — limiting dynamic steering capabilities to moving UAVs — measurable performance improvements were observed. The reference signal received power (RSRP) and signal to interference noise ratio (SINR) measurements showed consistent gains ranging from 2.5 dB to 4 dB across both frequency bands. These results validate the feasibility of RIS deployment for:

- ● **Beam Directionality:** Directing energy toward aerial targets;
- ● **Signal Enhancement:** Boosting desired signal strength;
- ● **Interference Mitigation:** Suppressing co-channel interference.

## VI. FUTURE TECHNOLOGY TRENDS

Although the solutions proposed in this paper address UAV signal coverage issues to some extent, several areas still require further research and optimization.

### A. Variability of the Atmospheric Ducting Effect

The atmospheric ducting effect exhibits significant variability due to its strong dependence on meteorological conditions, complicating the development of a universal model. To address this, researchers must conduct extensive experiments and data collection to deepen understanding of ducting effects across diverse environmental contexts. Concurrently, adaptive compensation techniques and real-time adjustment mechanisms are essential to maintain stable signal propagation under dynamic atmospheric conditions.

### B. Coordinated Optimization of Multi-layer Networks

A critical challenge in multi-layer network architectures lies in achieving coordinated optimization across heterogeneous network layers. Given the distinct characteristics and capabilities of network nodes at different layers, coordinating resource allocation and interference management becomes inherently complex. To resolve this, novel cooperative mechanisms and optimization algorithms are required to facilitate efficient inter-layer collaboration while accounting for layer-specific constraints.

### C. RIS-Based UAV Network Deployments

Addressing UAV signal coverage requires a multifaceted approach integrating RIS-based intelligent antenna systems, massive MIMO technologies, multi-layer network architectures, atmospheric ducting effect modeling, and security/privacy solutions. The overall performance of UAV communication systems can be significantly improved. While theoretical advancements provide foundational insights, future research must emphasize practical engineering implementation to bridge the gap between academic models and real-world deployments. This balance will ensure UAV communication systems achieve measurable performance improvements in operational environments.

### D. Engineering Implementation and Standardization Research

Current research predominantly focuses on theoretical investigations based on idealized assumptions. Future work must address system design under practical engineering constraints [137], including the following: (1) impacts of hardware implementation deviations in RIS and associated components; (2) effects of CSI measurement inaccuracies and quantization errors; (3) signaling overhead implications of CSI feedback mechanisms; (4) trade-offs between beamforming design complexity and performance optimization.

Furthermore, standardization is foundational for large-scale technological deployment. Enhanced standardization efforts are required for RIS-based UAV-ISAC systems, including: (1) standardized methodologies for channel characterization and modeling of atmospheric ducting effect; (2) optimized signal/channel design frameworks tailored for UAV-ISAC operational scenarios; (3) codebook design principles that incorporate mobility and channel variability constraints; (4) development of measurement and feedback signaling protocols with standardized process flows; and (5) interference management and coexistence mechanisms for heterogeneous networks. This standardization roadmap should integrate environmental considerations (e.g., atmospheric ducting phenomena) while ensuring compatibility with existing 3GPP specifications to facilitate seamless integration into future 6G network architectures.

## VII. CONCLUSION

Low-altitude coverage presents unique challenges compared to terrestrial network coverage. UAV networks, characterized by high mobility and distinct service distribution patterns, represent a typical application scenario for ISAC. The dual-layer heterogeneous topology of UAV-ISAC networks requires continuous 3D coverage, posing challenges for multi-layer network coexistence. Additionally, the distinct characteristics of UAV channels—such as the predominance of LoS paths—complicate the control of cell coverage areas and may exacerbate co-channel interference between neighboring cells. The up-tilted antenna configuration required for UAVs, in contrast to the down-tilted antenna configuration used in conventional terrestrial network coverage, can also intensify atmospheric ducting effects, particularly in TDD systems. This may lead to long-distance cross-link interference between cells.

To address these challenges, sub-6 GHz frequencies are suitable for supporting small-scale UAV low-altitude coverage through the deployment of new BSs. However, for large-scale UAV low-altitude coverage, new mid-band and mmWave frequency bands offer greater potential. RISs, with their advantages of low cost, low power consumption, and low complexity, hold promise as a key enabling technology for achieving seamless low-altitude coverage across these frequency bands. Consequently, RISs are poised to become a fundamental component of future wireless networks. A field trial was conducted to validate the RIS-based low-altitude coverage solution for UAVs, with test results demonstrating that RISs can effectively support such coverage.

Future research should focus on addressing the emerging challenges of low-altitude UAV coverage and exploring RIS-based solutions for continuous coverage in new mid-band and mmWave bands. Furthermore, efforts should prioritize accelerating the commercialization of non-standardized RIS-based products in the 5G-A phase, while advocating for RIS standardization in 6G. These steps will enable large-scale RIS deployment and unlock the full potential of UAV communication systems.

## REFERENCES

[1] A. Fotouhi et al., "Survey on UAV cellular communications: Practical aspects, standardization advancements, regulation, and security challenges," IEEE Commun. Surveys Tuts., vol. 21, no. 4, pp. 3417–3442, 4th Quart., 2019.

[2] M. Hassanalian and A. Abdelkefi, "Classifications, applications, and design challenges of drones: A review," Prog. Aerosp. Sci., vol. 91, pp. 99–131, May 2017.

[3] Y. Zeng, Q. Wu, and R. Zhang, "Accessing from the sky: A tutorial on UAV communications for 5G and beyond," Proc. IEEE, vol. 107, no. 12, pp. 2327–2375, Dec. 2019.

[4] M. Mozaffari, W. Saad, M. Bennis, Y.-H. Nam, and M. Debbah, "A tutorial on UAV for wireless networks: Applications, challenges, and open problems," IEEE Commun. Surveys Tuts., vol. 21, no. 3, pp. 2334–2360, 3rd Quart., 2019.

[5] Y. Zeng, I. Guvenc, R. Zhang, G. Geraci, and D. W. Matolak, UAV Communications for 5G and Beyond. Hoboken, NJ, USA: Wiley, 2020.

[6] M. Mozaffari, W. Saad, M. Bennis, Y.-H. Nam, and M. Debbah, "A tutorial on UAV for wireless networks: Applications, challenges, and open problems," IEEE Commun. Surveys Tuts., vol. 21, no. 3, pp. 2334–2360, 3rd Quart., 2019.

[7] S. A. H. Mohsan, M. A. Khan, F. Noor, I. Ullah, and M. H. Alsharif, "Towards the unmanned aerial vehicles (UAV): A comprehensive review," Drones, vol. 6, no. 6, p. 147, Jun. 2022.

[8] P. S. Bithas, E. T. Michailidis, N. Nomikos, D. Vouyioukas, and A. G. Kanatas, "Asurvey on machine-learning techniques for UAV-based communications," Sensors, vol. 19, no. 23, p. 5170, Nov. 2019.

[9] I. Shayea, P. Dushi, M. Banafaa, R. A. Rashid, S. Ali, M. A. Sarijari, Y. I. Daradkeh, and H. Mohamad, "Handover management for drones in future mobile networks—A survey," Sensors, vol. 22, no. 17, p. 6424, Aug. 2022.

[10] M. K. Banafaa et al., "A Comprehensive Survey on 5G-and-Beyond Networks With UAV: Applications, Emerging Technologies, Regulatory Aspects, Research Trends and Challenges," in IEEE Access, vol. 12, pp. 7786-7826, 2024, doi: 10.1109/ACCESS.2023.3349208.

[11] Y. Zeng, Q. Wu, and R. Zhang, "Accessing from the sky: A tutorial on UAV communications for 5G and beyond," Proc. IEEE, vol. 107, no. 12, pp. 2327–2375, Dec. 2019.

[12] ITU, "Framework and overall objectives of the future development of IMT for 2030 and beyond," Step. 2023.

[13] X. Wang, Z. Fei, J. A. Zhang, J. Huang and J. Yuan, "Constrained utility maximization in dual-functional radar-communication multi-UAV networks", IEEE Trans. Commun., vol. 69, no. 4, pp. 2660-2672, Apr. 2021.

[14] P. Kumari, N. J. Myers and R. W. Heath, "Adaptive and fast combined waveform-beamforming design for mmWave automotive joint communication-radar", IEEE J. Sel. Topics Signal Process., vol. 15, no. 4, pp. 996-1012, Jun. 2021.

[15] F. Liu et al., "Integrated sensing and communications: Toward dual-functional wireless networks for 6G and beyond," IEEE J. Sel. Areas Commun., vol. 40, no. 6, pp. 1728–1767, Jun. 2022.

[16] F. Liu, C. Masouros, A. Petropulu, H. Griffiths, and L. Hanzo, "Joint radar and communication design: Applications, state-of-the-art, and the road ahead," IEEE Trans. Commun., vol. 68, no. 6, pp. 3834–3862, Jun. 2020.

[17] J. A. Zhang et al., "An overview of signal processing techniques for joint communication and radar sensing," IEEE J. Sel. Topics Signal Process., vol. 15, no. 6, pp. 1295–1315, Nov. 2021.

[18] S. Depatla, C. R. Karanam, and Y. Mostofi, "Robotic through-wall imaging: Radio-frequency imaging possibilities with unmanned vehicles," IEEE Antennas Propag. Mag., vol. 59, no. 5, pp. 47–60, Oct. 2017.

[19] K. Meng et al., "UAV-enabled integrated sensing and communication: Opportunities and challenges," IEEE Wireless Commun., vol. 31, no. 2, pp. 97–104, Apr. 2024.

[20] J. Mu, R. Zhang, Y. Cui, N. Gao, and X. Jing, "UAV meets integrated sensing and communication: Challenges and future directions," IEEE Commun. Mag., vol. 61, no. 5, pp. 62–67, May 2023.

[21] Y. Li, X. Yuan, Y. Hu, J. Yang, and A. Schmeink, "Optimal UAV trajectory design for moving users in integrated sensing and communications networks," IEEE Trans. Intell. Transp. Syst., vol. 24, no. 12, pp. 15113–15130, Dec. 2023.

[22] K. Meng et al., "UAV-enabled integrated sensing and communication: Opportunities and challenges", IEEE Wireless Commun., Apr. 2023.

[23] X. Wang et al., "Constrained utility maximization in dual-functional radar-communication multi-UAV networks", IEEE Trans. Commun., vol. 69, no. 4, pp. 2660-2672, Apr. 2021.

[24] K. Meng, Q. Wu, S. Ma, W. Chen, K. Wang, J. Li, Throughput maximization for uav-enabled integrated periodic sensing and communication, IEEE Trans. Wirel. Commun. 22 (1) (2023) 671–687.

[25] Y. Liu, J. Zhao, M. Li, and Q. Wu, ''Intelligent reflecting surface aided MISO uplink communication network: Feasibility and power minimization for perfect and imperfect CSI,'' IEEE Trans. Commun., vol. 69, no. 3, pp. 1975–1989, Mar. 2021.

[26] Q. Wu et al., "A Comprehensive Overview on 5G-and-Beyond Networks With UAVs: From Communications to Sensing and Intelligence," in IEEE Journal on Selected Areas in Communications, vol. 39, no. 10, pp. 2912-2945, Oct. 2021, doi: 10.1109/JSAC.2021.3088681.

[27] W. Hurst, S. Evmorfos, A. Petropulu, and Y. Mostfoi. "Uncrewed Vehicles in 6G Networks: A Unifying Treatment of Problems, Formulations, and Tools," arXiv, 2024, doi: 10.48550/arXiv.2404.14738.

[28] M. Mozaffari, W. Saad, M. Bennis, Y.-H. Nam and M. Debbah, "A Tutorial on UAVs for Wireless Networks: Applications, Challenges, and Open Problems," in IEEE Communications Surveys & Tutorials, vol. 21, no. 3, pp. 2334-2360, 2019. doi: 10.1109/COMST.2019.2902862.

[29] B. Li, Z. Fei, and Y. Zhang, "UAV communications for 5G and beyond: Recent advances and future trends," IEEE Internet Things J., vol. 6, no. 2, pp. 2241–2263, Apr. 2019.

[30] J. Luo, Z. Wang, M. Xia, L. Wu, Y. Tian, and Y. Chen, "Path planning for UAV communication networks: Related technologies, solutions, and opportunities," ACM Comput. Surv., vol. 55, no. 9, pp. 1–37, Jan. 2023, doi: 10.1145/3560261.

[31] Y. Zeng, Q. Wu, and R. Zhang, "Accessing from the sky: A tutorial on UAV communications for 5G and beyond," Proc. IEEE, vol. 107, no. 12, pp. 2327–2375, Dec. 2019.

[32] M. Mozaffari, X. Lin, and S. Hayes, "Toward 6G with connected sky: UAVs and beyond," IEEE Commun. Mag., vol. 59, no. 12, pp. 74–80, Dec. 2021.

[33] C. Stöcker, R. Bennett, F. Nex, et al. "Review of the current state of UAV regulations," Remote Sens., vol. 9, no. 5, p. 459, 2017.

[34] M. Mozaffari, W. Saad, M. Bennis et al. ''A tutorial on UAV for wireless networks: Applications, challenges, and open problems,'' IEEE Commun. Surveys Tuts., vol. 21, no. 3, pp. 2334–2360, 3rd Quart., 2019.

[35] M. Shafi et al., "5G: A tutorial overview of standards, trials, challenges, deployment, and practice," IEEE J. Sel. Areas Commun., vol. 35, no. 6, pp. 1201–1221, Jun. 2017.

[36] N. Varshney and S. De, "Design Optimization for UAV Aided Sustainable 3D Wireless Communication at mmWaves," in IEEE Transactions on Vehicular Technology, vol. 72, no. 3, pp. 3274-3287, March 2023, doi: 10.1109/TVT.2022.3215908.

[37] S. Kang et al., "Coexistence of UAVs and Terrestrial Users in Millimeter-Wave Urban Networks," 2022 IEEE Globecom Workshops (GC Wkshps), Rio de Janeiro, Brazil, 2022, pp. 1158-1163, doi: 10.1109/GCWkshps56602.2022.10008629.

[38] Q. Wu et al., "A Comprehensive Overview on 5G-and-Beyond Networks With UAV: From Communications to Sensing and

Intelligence," in IEEE Journal on Selected Areas in Communications, vol. 39, no. 10, pp. 2912-2945, Oct. 2021, doi: 10.1109/JSAC.2021.3088681.

[39] A. Azari, F. Ghavimi, M. Ozger *et al.* " Machine learning assisted handover and resource management for cellular connected drones," in Proc. IEEE 91st Veh. Technol. Conf. (VTC-Spring), May 2020, pp. 1–7.

[40] S. Suman, S. Kumar and S. De, "Path Loss Model for UAV-Assisted RFET," in IEEE Communications Letters, vol. 22, no. 10, pp. 2048-2051, Oct. 2018, doi: 10.1109/LCOMM.2018.2863389.

[41] S. Sun *et al.*, "Propagation Path Loss Models for 5G Urban Micro- and Macro-Cellular Scenarios," 2016 IEEE 83rd Vehicular Technology Conference (VTC Spring), Nanjing, China, 2016, pp. 1-6, doi: 10.1109/VTCSpring.2016.7504435.

[42] DINC E, AKAN O B. Beyond-line-of-sight communications with ducting layer[J]. IEEE Communications Magazine, 2014, 52(10): 37-43.

[43] Study on remote interference management for NR[R]. 3GPP TR 38.866, 2018.

[44] F. Liu, J. Pan, X. Zhou and G. Y. Li, "Atmospheric Ducting Effect in Wireless Communications: Challenges and Opportunities," in Journal of Communications and Information Networks, vol. 6, no. 2, pp. 101-109, June 2021, doi: 10.23919/JCIN.2021.9475120.

[45] DINC E, AKAN O B. Beyond-line-of-sight communications with ducting layer[J]. IEEE Communications Magazine, 2014, 52(10): 37-43.

[46] LEVY M. Parabolic equation methods for electromagnetic wave propagation[M]. London: Institution of Electrical Engineers, 2000.

[47] M. Hu, K. Kang, T. Liu *et al.* "Cellular-Connected UAV: Recent Advances and Future Trends," 2024 IEEE 10th International Symposium on Microwave, Antenna, Propagation and EMC Technologies for Wireless Communications (MAPE), Guangzhou, China, 2024, pp. 1-4, doi: 10.1109/MAPE62875.2024.10813977.

[48] Q. Wu, X. Zhou, and R. Schober, ''IRS-assisted wireless powered NOMA: Do we really need different phase shifts in DL and UL?'' IEEE Wireless Commun. Lett., vol. 10, no. 7, pp. 1493–1497, Jul. 2021.

[49] C. Chaccour, M. N. Soorki, W. Saad, M *et al.* ''Risk-based optimization of virtual reality over terahertz reconfigurable intelligent surfaces,'' in Proc. IEEE Int. Conf. Commun. (ICC), Jun. 2020, pp. 1–6.

[50] H. Lu,Y. Zeng, S. Jin *et al.* ''Aerial intelligent reflecting surface: Joint placement and passive beamforming design with 3D beam flattening,'' IEEE Trans. Wireless Commun., vol. 20, no. 7, pp. 4128 – 4143, Jul. 2021.

[51] Y. J. Zhao. "Reconfigurable intelligent surfaces for 6G: applications, challenges, and solutions," Front Inform Technol Electron Eng 24, 1669–1688 (2023). https://doi.org/10.1631/FITEE.2200666

[52] Y. J. Zhao, L. L. Dai, J. H. Zhang, *et al.* "6G Near-field Technologies White Paper," FuTURE Forum, Nanjing, China, Apr 2024. doi: 10.12142/FuTURE.202404002.

[53] Y. J. ZHAO. "Reconfigurable Intelligent Surface Constructing 6G Near-Field Networks," Mobile Communications, 2024,48(4): 2-11.

[54] D. Ma, M. Ding and M. Hassan, "Enhancing cellular communications for UAVs via intelligent reflective surface", 2020 IEEE Wirel. Commun. and Networking Conf. (WCNC), pp. 1-6, 2020.

[55] A. Fotouhi, H. Qiang, M. Ding, *et al.,* "Survey on UAV cellular communications: Practical aspects standardization advancements regulation and security challenges", IEEE Commun. Surveys Tutorials, vol. 21, no. 4, pp. 3417-3442, 2019.

[56] Y. Zeng, Q. Wu, and R. Zhang, "Accessing from the sky: A tutorial on UAV communications for 5G and beyond," Proc. IEEE, vol. 107, no. 12, pp. 2327–2375, Dec. 2019.

[57] Y. Zeng, J. Lyu, and R. Zhang, "Cellular-connected UAV: Potential, challenges, and promising technologies," IEEE Wireless Commun., vol. 26, no. 1, pp. 120–127, Feb. 2019.

[58] W. Mei, Q. Wu, and R. Zhang, "Cellular-connected UAV: Uplink association, power control and interference coordination," IEEE Trans. Wireless Commun., vol. 18, no. 11, pp. 5380–5393, Nov. 2019.

[59] S. Zhang, Y. Zeng, and R. Zhang, "Cellular-enabled UAV communication: A connectivity-constrained trajectory optimization perspective," IEEE Trans. Commun., vol. 67, no. 3, pp. 2580–2604, Mar. 2019.

[60] S. D. Muruganathan *et al.*, "An overview of 3GPP release-15 study on enhanced LTE support for connected drones," 2018, arXiv:1805.00826. [Online]. Available: http://arxiv.org/abs/1805.00826

[61] X. Lin *et al.*, "The sky is not the limit: LTE for unmanned aerial vehicles," IEEE Commun. Mag., vol. 56, no. 4, pp. 204–210, Apr. 2018.

[62] S. Homayouni, M. Paier, C. Benischek, *et al.* ''On the feasibility of cellular-connected drones in existing 4G/5G networks: Field trials,'' in Proc. IEEE 4th 5G World Forum (5GWF), Oct. 2021, pp. 287–292.

[63] S. Homayouni, M. Paier, C. Benischek, *et al.* "Field trials and design insights of cellular-connected drones," in Proc. IEEE 94th Veh. Technol. Conf. (VTC-Fall), Sep. 2021, pp. 1–6.

[64] R. Amer, W. Saad, B. Galkin, *et al.* "Performance analysis of mobile cellular-connected drones under practical antenna configurations," in Proc. IEEE Int. Conf. Commun. (ICC), Jun. 2020, pp. 1–7.

[65] K. Sun, J. Yang, C.K. *et al.,* "Channel Charting for UAV Navigation in RIS-Assisted ISAC Systems," Submitted to IEEE Journal of Selected Topics in Electromagnetics. Antennas and Propagation, 2025.

[66] M. Jian *et al.,* "Reconfigurable intelligent surfaces for wireless communications: Overview of hardware designs, channel models, and estimation techniques," in Intelligent and Converged Networks, vol. 3, no. 1, pp. 1-32, March 2022, doi: 10.23919/ICN.2022.0005.

[67] L. Dai *et al.,* "Reconfigurable Intelligent Surface-Based Wireless Communications: Antenna Design, Prototyping, and Experimental Results," in IEEE Access, vol. 8, pp. 45913-45923, 2020, doi: 10.1109/ACCESS.2020.2977772.

[68] C. DAndrea, A. Garcia-Rodriguez, G. Geraci, *et al.* "Analysis of UAV communications in cell-free massive MIMO systems", IEEE Open Journal of the Commun. Society, vol. 1, pp. 133-147, 2020.

[69] H. Zhang, N Shlezinger, F. Guidi, *et al.,* "Beam Focusing for Near-field Multi-user MIMO Communications," IEEE Transactions on Wireless Communications, vol. 21, no. 9, pp. 7476-7490, September 2022.

[70] H. Zhang, N. Shlezinger, F. Guidi, *et al.,* "6G Wireless Communications: From Far-field Beam-steering to Near-field Beam-focusing, " IEEE Communications Magazine, vol. 61, no. 4, pp. 72-77, April 2023.

[71] J. Durnin. "Exact solutions for nondiffracting beams. I. The scalar theory," Journal of the Optical Society of America 1987, 4(4): 651-654.

[72] M. V. Berry and N. L. Balazs, "Nonspreading wave packets", Amer. J. Phys., vol. 47, no. 3, pp. 264-267, Mar. 1979.

[73] J. Durnin, J. J. Miceli and J. H. Eberly, "Diffraction-free beams", Phys. Rev. Lett., vol. 58, pp. 1499-1501, Apr. 1987.

[74] J. Durnin, "Exact solutions for nondiffracting beams. I. The scalar theory", J. Opt. Soc. Amer. A Opt. Image Sci., vol. 4, no. 4, pp. 651-654, 1987.

[75] Y.-Y. Shi, W. Tong, L. You-Wen, *et al.* "Control of self-bending Airy beams", Acta Photonica Sinica, vol. 42, pp. 1401-1407, Dec. 2013.

[76] J. Durnin, J. J. Miceli, J. H. Eberly. Diffraction-free beams. Phy Rev Lett 1987; 58: 1499–1501.

[77] Z. Bouchal, J. Wagner, M. Chlup . Self-reconstruction of a distorted nondiffracting beam. Opt Commun 1998; 151: 207–211.

[78] J. Chen, J. Ng, Z. F. Lin *et al.,* Optical pulling force. Nat Photon 2011; 5: 531–534.

[79] A. Dogariu, S. Sukhov, J. Jose Sáenz. Optically induced 'negative forces'. Nat Photon 2013; 7: 24–27.

[80] C. Zhang, W. Zhang, W. Wang, L. Yang and W. Zhang, "Research Challenges and Opportunities of UAV Millimeter-Wave Communications," in IEEE Wireless Communications, vol. 26, no. 1, pp. 58-62, February 2019, doi: 10.1109/MWC.2018.1800214.

[81] Y. J. Zhao, L. L. Dai, J. H. Zhang *et al.,* "Near-field communications: characteristics, technologies, and engineering," Front Inform Technol Electron Eng 25, 1580 – 1626 (2024). https://doi.org/10.1631/FITEE.2400576.

[82] M. Jian, A. Tang, Y. Chen and Y. Zhao, "Fractional Fourier Transformation Based XL-MIMO Near-Field Channel Analysis," 2024 IEEE 25th International Workshop on Signal Processing Advances in Wireless Communications (SPAWC), Lucca, Italy, 2024, pp. 221-225, doi: 10.1109/SPAWC60668.2024.10694136.

[83] M. Eskandari and A. V. Savkin, "SLAPS: Simultaneous Localization and Phase Shift for a RIS-Equipped UAV in 5G/6G Wireless Communication Networks," in IEEE Transactions on Intelligent Vehicles, vol. 8, no. 12, pp. 4722-4733, Dec. 2023, doi: 10.1109/TIV.2023.3298607.

[84] Z. Du *et al.*, "Integrated Sensing and Communications for V2I Networks: Dynamic Predictive Beamforming for Extended Vehicle Targets," in IEEE Transactions on Wireless Communications, vol. 22, no. 6, pp. 3612-3627, June 2023, doi: 10.1109/TWC.2022.3219890.

[85] Q. T. Ngo, K. T. Phan, A. Mahmood, and W. Xiang, "DRL-based secure beamforming for hybrid-RIS aided satellite downlink communications," in Proc. IEEE COMNETSAT, Nov. 2022, pp. 432–437.

[86] X. Liu, Y. Liu, and Y. Chen, "Machine learning empowered trajectory and passive beamforming design in UAV-RIS wireless networks," IEEE J. Sel. Areas Commun., vol. 39, no. 7, pp. 2042–2055, Jul. 2021.

[87] L. Wang, K. Wang, C. Pan, and N. Aslam, "Joint trajectory and passive beamforming design for intelligent reflecting surface-aided UAV communications: A deep reinforcement learning approach," IEEE Trans. Mob. Comput., vol. 22, no. 11, pp. 6543–6553, Nov. 2023.

[88] P. Saikia, S. Pala, K. Singh, S. K. Singh, and W.-J. Huang, "Proximal policy optimization for RIS-assisted full duplex 6G-V2X communications," IEEE Trans. Intell. Veh., pp. 1–16, May 2023.

[89] Y. J. Zhao, L. L. Dai, J. H. Zhang, "Near-field communications: characteristics, technologies, and engineering," Front Inform Technol Electron Eng 25, 1580 — 1626 (2024). https://doi.org/10.1631/FITEE.2400576.

[90] X. Wang, Z. Fei, J. Huang and H. Yu, "Joint Waveform and Discrete Phase Shift Design for RIS-Assisted Integrated Sensing and Communication System Under Cramer-Rao Bound Constraint", IEEE Transactions on Vehicular Technology, vol. 71, no. 1, pp. 1004-1009, Jan. 2022.

[91] Y. He, Y. Cai, H. Mao and G. Yu, "RIS-Assisted Communication Radar Coexistence: Joint Beamforming Design and Analysis", IEEE Journal on Selected Areas in Communications, vol. 40, no. 7, pp. 2131-2145, July 2022.

[92] H. Luo, R. Liu, M. Li, Y. Liu and Q. Liu, "Joint Beamforming Design for RIS-Assisted Integrated Sensing and Communication Systems", IEEE Transactions on Vehicular Technology, vol. 71, no. 12, pp. 13393-13397, Dec. 2022.

[93] M. Cui, L. Dai, Z. Wang, S. Zhou, and N. Ge, "Near-field rainbow: Wideband beam training for XL-MIMO," IEEE Trans. Wireless Commun., vol. 22, no. 6, pp. 3899–3912, Jun. 2023.

[94] W. Liu, H. Ren, C. Pan and J. Wang, "Deep Learning Based Beam Training for Extremely Large-Scale Massive MIMO in Near-Field Domain," in IEEE Communications Letters, vol. 27, no. 1, pp. 170-174, Jan. 2023.

[95] W. Liu, C. Pan, H. Ren, F. Shu, S. Jin and J. Wang, "Low-Overhead Beam Training Scheme for Extremely Large-Scale RIS in Near Field," in IEEE Transactions on Communications, vol. 71, no. 8, pp. 4924-4940, Aug. 2023.

[96] Y. J. Zhao. "Cascaded channel decoupling based solution for RIS reconfiguration matrix," Intelligent and Converged Networks, 2024, 5(1): 19-27.

[97] Q He, J Li, Z Wang, et al., "Comparative Study of the 60 GHz and 118 GHz Oxygen Absorption Bands for Sounding Sea Surface Barometric Pressure," Remote Sensing. 2022; 14(9):2260. https://doi.org/10.3390/rs14092260.

[98] K. B. Cooper et al., "Atmospheric Humidity Sounding Using Differential Absorption Radar Near 183 GHz," in IEEE Geoscience and Remote Sensing Letters, vol. 15, no. 2, pp. 163-167, Feb. 2018, doi: 10.1109/LGRS.2017.2776078.

[99] H. Q. Ngo, A. Ashikhmin, H. Yang, E. G. Larsson and T. L. Marzetta, "Cell-free massive MIMO versus small cells," IEEE Trans. Wireless Commun., vol. 16, no. 3, pp. 1834-1850, Mar. 2017.

[100] G. Interdonato, E. Björnson, H. Quoc Ngo, et al., "Ubiquitous cell-free Massive MIMO communications," Wireless Com Network 2019, 197 (2019). https://doi.org/10.1186/s13638-019-1507-0.

[101] S. Elhoshy et al., "A dimensioning framework for indoor DAS LTE networks, " in Proc. Int. Conf. Sel. Topics Mobile Wireless Netw. (MoWNeT), 2016, pp. 1–8.

[102] R. Irmer et al., "Coordinated multipoint: Concepts, performance, and field trial results," IEEE Commun. Mag., vol. 49, no. 2, pp. 102–111, Feb. 2011.

[103] H. Q. Xu, Y. J. Zhao, L. M. Mo, C. Huang and B. Sun, "Inter-cell antenna calibration for coherent joint transmission in TDD system," 2012 IEEE Globecom Workshops, Anaheim, CA, USA, 2012, pp. 297-301, doi: 10.1109/GLOCOMW.2012.6477586.

[104] Zhao, Yajun. "RIS assisted wireless networks: Collaborative regulation, deployment mode and field testing." IET Communications (2024).

[105] Z. Zhou, X. Gao, J. Fang, and Z. Chen, "Spherical wave channel and analysis for large linear array in LoS conditions, " in Proc. IEEE Globecom Workshops2015, pp. 1–6.

[106] Zhao YJ, Dai LL, Zhang JH, et al., 2024. 6G Near-Field Technologies White Paper. FuTURE Forum, Nanjing, China. https://doi.org/10.12142/FuTURE.202404002.

[107] A. Yaghjian, "An overview of near-field antenna measurements," IEEE Trans Antennas Propag, 34(1):30-45. 1986. https://doi.org/10.1109/TAP.1986.1143727.

[108] Selvan KT, Janaswamy R. Fraunhofer and Fresnel distances: unified derivation for aperture antennas. IEEE Antennas Propag Mag, 59(4):12-15, 2017. https://doi.org/10.1109/MAP.2017.2706648.

[109] Ramezani P, Kosasih A, Irshad A, et al., 2023. Exploiting the depth and angular domains for massive near-field spatial multiplexing. IEEE BITS Inf Theory Mag, 3(1):14-26. https://doi.org/10.1109/MBITS.2023.3322670.

[110] Y. J. Zhao. "Reconfigurable Intelligent Surface Constructing 6G Near-field Networks," Authorea Preprints. 2024 Jan 29.

[111] M. A. Jasim, H. Shakhatreh, N. Siasi, A. H. Sawalmeh, A. Aldalbahi, and A. Al-Fuqaha, "A survey on spectrum management for unmanned aerial vehicles (UAV)," IEEE Access, vol. 10, pp. 11443–11499, 2022.

[112] Y. J.Zhao, M. N. Jian. "Applications and challenges of reconfigurable intelligent surface for 6G networks. Radio Communications Technology," 6, 1–16 (2021)

[113] Y. J.Zhao, X. Lv. "Network coexistence analysis of RIS-assisted wireless communications," IEEE Access 10, 63442–63454 (2022).

[114] J. C. Liang ,L. Zhang, ,Z.Luo, et al., "A filtering reconfigurable intelligent surface for interference-free wireless communications," Nature Communications, 15(1), p.3838. 2024.

[115] Liang J C., Zhang L., Luo Z., et al., "A filtering reconfigurable intelligent surface for interference-free wireless communications," Nature Communications, 15(1), 3838. (2024).

[116] D. Gürgünoğlu, E. Björnson and G. Fodor, "Combating Inter-Operator Pilot Contamination in Reconfigurable Intelligent Surfaces Assisted Multi-Operator Networks," in IEEE Transactions on Communications, doi: 10.1109/TCOMM.2024.3390095.

[117] Z. Xiao, L. Zhu, Y. Liu, et al., "A survey on millimeter-wave beamforming enabled UAV communications and networking," IEEE Commun. Surveys Tuts., vol. 24, no. 1, pp. 557–610, 1st Quart., 2022.

[118] S. K. Khan, U. Naseem, H. Siraj, et al., ''The role of unmanned aerial vehicles andmmWave in 5G: Recent advances and challenges,'' Trans. Emerg. Telecommun. Technol., vol. 32, no. 7, p. e4241, Jul. 2021.

[119] C. Zhang, W. Zhang, W. Wang et al., "Research challenges and opportunities of UAV millimeter-wave communications," IEEE Wireless Commun., vol. 26, no. 1, pp. 58–62, Feb. 2019.

[120] Johnson, Richard. Antenna Engineering Handbook 2nd. New York, NY: McGraw-Hill, Inc. 1984: 1-12. ISBN 0-07-032291-0.

[121] H. T. Friis, "A Note on a Simple Transmission Formula," in Proceedings of the IRE, vol. 34, no. 5, pp. 254-256, May 1946, doi: 10.1109/JRPROC.1946.234568.

[122] X. Yang, Z. Wei, Y. Liu et al., "RIS-Assisted Cooperative Multicell ISAC Systems: A Multi-User and Multi-Target Case," in IEEE Transactions on Wireless Communications, vol. 23, no. 8, pp. 8683-8699, Aug. 2024, doi: 10.1109/TWC.2024.3353336.

[123] K. Meng, Q. Wu, S. Ma et al., "Throughput Maximization for UAV-Enabled Integrated Periodic Sensing and Communication," IEEE Transactions on Wireless Communications, vol. 22, no. 1, pp. 671-687, Jan. 2023.

[124] "3rd Generation Partnership Project Technical Specification Group Radio Access Network", 3GPP Tech. Rep., March 2017.

[125] "TR 36.777: Technical Specification Group Radio Access Network; Study on Enhanced LTE Support for Aerial Vehicles (Release 15)", 3GPP Tech. Rep., 2018.

[126] "TS 36.331: Technical specification group radio access network; evolved universal terrestrial radio access (e-utra); radio resource control (rrc); protocol specification (release 16)", 3GPP Tech. Rep., 2020.

[127] S. D. Muruganathan et al., "An overview of 3GPP release-15 study on enhanced LTE support for connected drones," 2018, arXiv:1805.00826. [Online]. Available: http://arxiv.org/abs/1805.00826

[128] Accessed: Aug. 2020. [Online]. Available: https://www.3gpp.org/uas-uav

[129] Mohsan SAH, Khan MA, Noor F, Ullah I, Alsharif MH. Towards the Unmanned Aerial Vehicles (UAVs): A Comprehensive Review. Drones. 2022; 6(6):147. https://doi.org/10.3390/drones6060147

[130] Functional architecture for unmanned aerial vehicles and unmanned aerial vehicle controllers using IMT-2020 networks (2017). ITU-T, Geneva

[131] Use cases and spectrum considerations for UAS (unmanned aircraft systems) (2018). ETSI, Sophia Antipolis, France, Rep. 103 373

[132] Accessed: Aug. 2020. [Online]. Available: https://www.comsoc.org/about/committees/emerging-technologies-initiatives/aerialcommunications

[133] Accessed: Aug. 2020. [Online]. Available: https://ieeexplore.ieee.org/stamp/stamp.jsp?arnumber=9097766

[134] Accessed: Aug. 2020. [Online]. Available: https://aceti.committees.comsoc.org/initiatives/

[135] 3GPP. "Technical Specification Group Radio Access Network; Study on remote interference management for NR (Release 16)", 3rd Generation Partnership Project (3GPP) TR 38.866 V16.0.0, Dec. 2018.

[136] 3GPP. RP-240799 Study on channel modelling for Integrated Sensing And Communication (ISAC) for NR, 3GPP TSG RAN Meeting #105, Melbourne, Australia, September 9-12, 2024.

[137] X. Zhao, M. Jian, Y. Chen, Y. Zhao and L. Mu, "Reconfigurable Intelligent Surfaces for 6G: Engineering Challenges and the Road Ahead," in Intelligent and Converged Networks, vol. 6, no. 1, pp. 53-81, March 2025, doi: 10.23919/ICN.2025.0004.

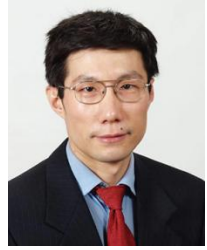

**Yifei Yuan** (Fellow, IEEE) received the bachelor's and master's degrees from Tsinghua University, Beijing, China, in 1993 and 1996, respectively, and the Ph.D. degree from Carnegie Mellon University, Pittsburgh, PA, USA, in 2000. He was with Lucent Technologies Bell Labs, Murray Hill, NJ, USA, from 2000 to 2008. From 2008 to 2020, he was a Technical Director and a Chief Engineer with ZTE Corporation, Shenzhen, China, responsible for standards and research of LTE-Advanced and 5G. He has been the Chief Expert with China Mobile Research Institute, Beijing, since 2020, responsible for 6G. He has extensive publications, including 12 books on key air interface technologies of 4G, 5G, and 6G mobile communications. He has more than 60 granted U.S. patents.

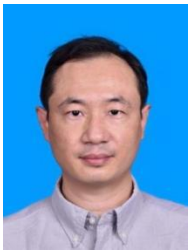

**Zhao Yajun** (Member, IEEE) holds Bachelor's, Master's, and Doctoral degrees. Since 2010, he has assumed the role of Chief Engineer within the Wireless and Computing Product R&D Institute at ZTE Corp. Prior to this, he contributed to wireless technology research within the Wireless Research Department at Huawei. Currently, his primary focus centers on 5G standardization technology and the advancement of future mobile communication technology, particularly 6G. His research pursuits encompass a broad spectrum, including reconfigurable intelligent surface (RIS), spectrum sharing, flexible duplex, CoMP, and interference mitigation. He has played an instrumental role in founding the RIS TECH Alliance (RISTA) and currently holds the position of Deputy Secretary General within the organization. Additionally, he is a founding member of the RIS task group under the purview of the China IMT-2030 (6G) Promotion Group, where he serves as the Deputy Leader. Up to now, he has more than 200 granted patents related to 4G LTE and 5G NR mobile communication technologies, and more than 20 of them have been adopted into the 4G/5G standard and become the standard essential patents.

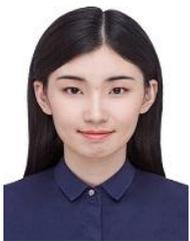

**Mengnan Jian** (Member, IEEE) received the B.S. degree in information engineering from the Beijing Institute of Technology, Beijing, China, in 2016 and the M.S. degree from Tsinghua University, Beijing, China, in 2019. She is currently an senior ngineer in ZTE cooperation. Her research interests include massive MIMO, holographic MIMO, distributed MIMO and reconfigurable intelligent surface.